%% file: article.tex
\documentclass[
    amsmath,
	amssymb,
	a4paper,
	aps,		% society option (aps, aip)
	pre,		% society's journal style (prl, apl, ..)
	onecolumn,	% layout option (preprint, reprint, twocolumn)
	showpacs,
	floatfix,
	notitlepage,
	superscriptaddress
]{revtex4-1}
    
\usepackage[lining, semibold]{libertine}
\usepackage[libertine, slantedGreek, bigdelims, vvarbb]{newtxmath}

\usepackage{xcolor}
\usepackage{braket}
\usepackage{bm}
\usepackage{mathtools}
\usepackage{kbordermatrix}
	\renewcommand{\kbldelim}{(}% Left delimiter
	\renewcommand{\kbrdelim}{)}% Right delimiter

\usepackage{dcolumn}% Align table columns on decimal point
\usepackage{tabularx}
\usepackage{booktabs}
\usepackage{multirow}

\usepackage{import}
\usepackage[caption=false]{subfig}
\usepackage{graphicx}
\usepackage{svg}        
\usepackage{pgfplots}
\usepgfplotslibrary{colorbrewer}
\graphicspath{./figures/}

\definecolor{webgreen}{rgb}{0,.5,0}
\definecolor{webbrown}{rgb}{.6,0,0}
\definecolor{grigio}{rgb}{.85,.85,.85} 
\definecolor{RoyalBlue}{rgb}{0.0, 0.14, 0.4}
\definecolor{skyblue1}{rgb}{0.45,0.62,0.81}
\definecolor{skyblue2}{rgb}{0.2,0.39,0.64}
\definecolor{skyblue3}{rgb}{0.13,0.29,0.53}
\definecolor{scarlet1}{rgb}{0.93,0.16,0.16}
\definecolor{scarlet2}{rgb}{0.8,0,0}
\definecolor{scarlet3}{rgb}{0.64,0,0}

\usepackage[normalem]{ulem}
\usepackage{hyperref}
\hypersetup{%
    colorlinks=true, linktocpage=true, pdfstartpage=1, pdfstartview=FitV,%
    breaklinks=true, pdfpagemode=UseNone, pageanchor=true, pdfpagemode=UseOutlines,%
    plainpages=false, bookmarksnumbered, bookmarksopen=true, bookmarksopenlevel=1,%
    hypertexnames=true, pdfhighlight=/O,%nesting=true,%frenchlinks,%
    urlcolor=webbrown, linkcolor=RoyalBlue, citecolor=webgreen, %pagecolor=RoyalBlue,%
    pdftitle={Linear Stochastic Thermodynamics},%
    pdfauthor={Danilo Forastiere, Riccardo Rao, Massimiliano Esposito,}%
    pdfsubject={},%
    pdfkeywords={},%
    pdfcreator={pdfLaTeX},%
    pdfproducer={LaTeX REVTeX}%
  }

\input{./texfiles/mydefs}

\begin{document}
%\tableofcontents
\title{Linear Stochastic Thermodynamics}
\author{Danilo Forastiere}
\affiliation{Complex Systems and Statistical Mechanics, Department of Physics and Materials Science, University of Luxembourg, L-1511 Luxembourg}
\author{Riccardo Rao}
\affiliation{Simons Center for Systems Biology, School of Natural Sciences, Institute for Advanced Study, 08540 Princeton (NJ), U.S.A.}
\author{Massimiliano Esposito}
\affiliation{Complex Systems and Statistical Mechanics, Department of Physics and Materials Science, University of Luxembourg, L-1511 Luxembourg}
\date{\today}
\begin{abstract}
	\input{./texfiles/abstract}
\end{abstract}

% =================================================================
\maketitle
% =================================================================

\input{./texfiles/draft}
\bibliography{./bibliography}
\end{document}

%% file: texfiles/mydefs.tex
\DeclareMathOperator{\e}{\mathrm{e}} %Euler number
\renewcommand*{\i}{\mathrm{i}}
\newcommand*{\abs}[1]{\left\lvert{#1}\right\rvert}

\renewcommand*{\vec}[1]{\bm{#1}} 

\newcommand{\tr}{^{\mathsf{T}}}

\makeatletter
\renewcommand*{\d}%
{\@ifnextchar^{\DIfF}{\DIfF^{}}}
\def\DIfF^#1{%
  \mathop{\mathrm{\mathstrut d}}%
  \nolimits^{#1}\gobblespace
}
\def\gobblespace{%
  \futurelet\diffarg\opspace}
\def\opspace{%
  \let\DiffSpace\!%
  \ifx\diffarg(%
  \let\DiffSpace\relax
  \else
  \ifx\diffarg\[%
    \let\DiffSpace\relax
    \else
    \ifx\diffarg\{%
    \let\DiffSpace\relax
    \fi\fi\fi\DiffSpace
  }

\newcommand*{\pder}[2][]{\frac{\partial#1}{\partial#2}} 			
\providecommand*{\avg}[1]{\left\langle#1\right\rangle}

%% file: texfiles/abstract.tex
We study the thermodynamics of open systems weakly driven out-of-equilibrium by nonconservative and time-dependent forces using the linear regime of stochastic thermodynamics. 
We make use of conservation laws to identify the potential and nonconservative components of the forces. This allows us to formulate a unified near-equilibrium thermodynamics. 
For nonequilibrium steady states, we obtain an Onsager theory ensuring nonsingular response matrices that is consistent with phenomenological linear irreversible thermodynamics. 
For time-dependent driving protocols that do not produce nonconservative forces, we identify the equilibrium ensemble from which Green--Kubo relations are recovered. 
For arbitrary periodic drivings, the averaged entropy production (EP) is expressed as an independent sum over each driving frequency of non-negative contributions.
These contributions are bilinear in the nonconservative and conservative forces and involve a novel generalized Onsager matrix that is symmetric. 
%The EP for a given driving and its time-reversal are identical. 
In the most general case of arbitrary time-dependent drivings, we advance a novel decomposition of the EP rate into two non-negative contributions -- one solely due to nonconservative forces and the other solely due to deviation from the instantaneous steady-state -- directly implying a minimum entropy production principle close to equilibrium. 
This setting reveals the geometric structure of near-equilibrium thermodynamics and generalizes previous approaches to cases with nonconservative forces.

%%% Local Variables:
%%% mode: latex
%%% TeX-master: "../article"
%%% End:

%% file: texfiles/draft.tex
\section{Introduction}
\label{sec:intro}

The development of phenomenological irreversible thermodynamics in the first half of the twentieth century primarily relies on the concept of local equilibrium. Since macroscopic bulk systems are considered, equilibrium thermodynamics is assumed to hold within each volume element of the system.
Exchange processes of globally conserved quantities (e.g. energy and mass) between nearby volumes cause an irreversible entropy production. This fundamental quantity determines the amount of entropy which is irreversibly dissipated over time, and it can be evaluated by relying on a second assumption: the gradients of intensive quantities (e.g. temperature and chemical potential) are locally small enough to justify linearizing the currents of the conjugated extensive conserved quantities (e.g. energy flow for temperature gradients and particle flow for chemical potential gradients) \cite{prigogine1947etude, callen1948application, prigogine1949domaine, dgm1962nonequilibrium, glansdorff1971thermodynamic, callen1985thermodynamics}. A third important assumption is that the Onsager matrix resulting from that linearization is symmetric.    
Justifications of this fact rely on microreversibility and detailed balance, which holds at equilibrium. They follow two main directions. In the first, the internal dynamics of the system is not modeled but linear stochastic equations are assumed to describe the fluxes of extensive quantities~\cite{onsager1931reciprocalI, onsager1931reciprocalII, callen1952statistical, callen1952theorem, greene1952theorem,  wigner1954derivations}. 
The second approach is based on linear response theories, where the internal dynamics of a system initially described by an equilibrium distribution is weakly perturbed from the outside. The dynamical response of the system near equilibrium can then be expressed in terms of equilibrium correlators~\cite{kirkwood1946statistical, callen1951irreversibility, green1952markoff, green1954markoff, bergmann1955new, kubo1957statistical}. 

Stochastic thermodynamics (ST) is a more modern endeavor which formulates nonequilibrium thermodynamics for systems far from equilibrium in contact with reservoirs and described by stochastic dynamics~\cite{schnakenberg1976network,sekimoto2010stochastic, hill2005free, jiu1984stability, mou1986stochastic, jarzynski1997equilibrium, crooks1999entropy, seifert2005entropy, seifert2012stochastic, zhang2012stochastic, rao2018conservation, sekimoto2010stochastic, pigolotti2021}, and has therefore become the standard setting to model far from equilibrium mesoscopic systems. The scope of stochastic thermodynamics broadened with time: it started from describing the dynamics of the averages of thermodynamic observables~\cite{bergmann1955new, schnakenberg1976network,jiu1984stability, hill2005free} and arrived at a precise identification of their fluctuating analogs. The central assumption of local detailed balance (LDB) relates features of the noise induced by the reservoirs with the entropy change in the reservoir~\cite{bergmann1955new, esposito2012stochastic, maes2020local, falasco2021local}. This assumption can be justified microscopically by assuming that the reservoirs are only weakly displaced from equilibrium by the system \cite{esposito2009nonequilibrium, breuer2002theory}, and it is a key ingredient to study systems driven arbitrarily far from equilibrium. In fact, the LDB property allows to translate in the language of stochastic thermodynamics a major breakthrough in nonequilibrium statistical physics, namely fluctuation theorems~\cite{bochkov1977general, evans1993probability, gallavotti1994dynamical, gallavotti1996extension}. These latter provide a refinement of the second law \cite{jarzynski1997equilibrium} and enables to extend thermodynamics at the trajectory level~\cite{ kurchan1998fluctuation,  lebowitz1999gallavotti, maes1999fluctuation, crooks1999entropy}. By now, many experimental validations of ST are available~\cite{pekola2015towards, ciliberto2017experiments} (and references therein). 

%or on linear response around nonequilibrium steady states far from equilibrium \cite{prost2009generalized,seifert2010, baiesi2013response, vroylandt2018degree,owen2021universal}.
%
%Links to NESS were uncovered~\cite{raz2016mimicking, rotskoff2017mapping, busiello2018similarities}

To fix the terminology, let us consider a system in contact with two thermostats at temperatures $T_1(t)$ and $T_2(t)$, which may depend on time. For $T_1(t) \neq T_2(t)$, the system is pushed away from equilibrium by the nonconservative thermodynamic force $\mathcal{F}(t)=\frac{1}{T_1(t)}-\frac{1}{T_2(t)}$. If $\mathcal{F}(t)$ changes in time because of a driving protocol, this driving is said to be \emph{nonconservative}. In contrast, if $\mathcal{F}(t)=0$ at all times, both the system and the driving protocol are said to be \emph{detailed balanced}. 
%In that latter case, an instantaneous equilibrium distribution capable of restoring detailed balance exists at all times, but the system will not be able to relax to it the protocol is too fast. 
It may happen that a system does not feature any nonconservative force, for example if a single thermal reservoir constitutes the whole environment.
In this case, the driving is detailed balanced by construction, and we speak of an \emph{unconditionally detailed balance} driving and system.
Linear response theory provides a natural tool to explore the close-to-equilibrium regime of stochastic thermodynamics and establish connections with phenomenological irreversible thermodynamics. 
Many such studies have been carried out in the past, but most consider unconditionally detailed balance drivings (arbitrary~\cite{crooks2007measuring, feng2009far, sivak2012thermodynamic} or periodic~\cite{izumida2009onsager, brandner2015thermodynamics, proesmans2015onsager, proesmans2016linear, cleuren2020stochastic, proesmans2019general}) or nonequilibrium steady states \cite{lebowitz1999gallavotti, andrieux2004fluctuation}.
Few also described systems driven by time-dependent nonconservative forces \cite{brandner2016periodic, proesmans2019general}.
However, these descriptions lack a systematic procedure to decompose the driving into its conservative and nonconservative contributions, which is important since these two contributions generate very different kinds of responses.
A procedure to achieve this separation based on conservation laws is a quite recent achievement~\cite{polettini2016conservation, rao2018conservation}.
In this paper, we build on it to formulate a linear response theory of stochastic thermodynamics for arbitrary drivings that makes contact with classic results of irreversible thermodynamics. 
We develop our theory using Markov jump processes at the ensemble averaged level. 
For nonequilibrium steady states, macroscopic theory for steady-state transport is recovered, and a symmetric nonsingular Onsager matrix ensues by construction. 
For unconditionally detailed-balanced drivings, we show how to recover Green--Kubo relations for the response of state observables. 
For periodically driven systems, we introduce a novel frequency-resolved generalized Onsager matrix ~\eqref{eq:onsagerdef} that -- in contrast to previous descriptions -- is symmetric and  independent on the driving protocol. 
For arbitrary protocols, we show that the EPR decomposes into two positive-definite quadratic forms, Eq.~\eqref{eq:sstransient}. The first reduces to the total EPR in a NESS and contains the infinite-time response matrix of physical currents, \textit{i.e.} the Onsager matrix computed at steady state. The second gives the dissipation due to the lag of the dynamics with respect to the instantaneous steady state identified by the forces. The minimum entropy production principle follows immediately.

\paragraph*{Outline.} We start by reviewing stochastic thermodynamics of Markov jump processes using conservation laws, section~\ref{sec:stochasticthermo}. In section~\ref{sec:driving_reservoirs}, we then formulate a linear response theory for perturbations only acting on the intensive field that characterize the reservoirs. The generalisation to perturbations that also act on system quantities is presented in section~\ref{sec:generalprotocol}. The theory is illustrated on simple systems in Section~\ref{sec:examples} and conclusions are drawn in Section~\ref{sec:conclusions}.

\section{Stochastic thermodynamics and conservation laws}
\label{sec:stochasticthermo}

In this section we revisit the formulation of stochastic thermodynamics that makes use of conservation laws to discriminate between conservative and nonconservative driving forces. 

\subsection{Thermodynamics for Markov jump processes}

\begin{figure}[htb]
    \centering
    \def\svgwidth{0.5\textwidth}
      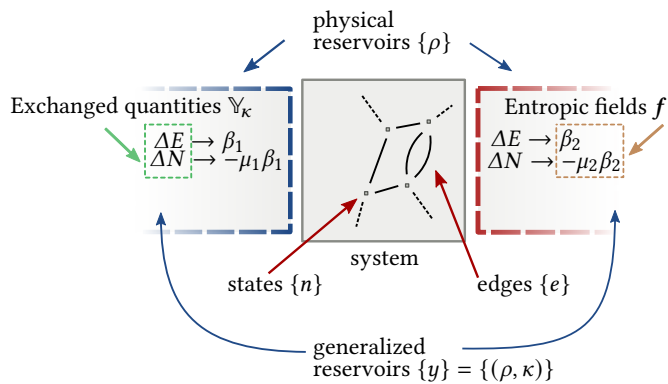
    \caption{Schematic representation of a system coupled to two reservoirs. Both can exchange energy and particles with the system through various transition mechanisms.}
    \label{fig:schematics}
\end{figure}

We consider a system composed of a set $\mathcal{V}=\{n\}$ of mesoscopic states to which one assigns a set $\mathcal{K}=\{\kappa\}$ of different extensive quantities (e.g. energy, particle numbers), denoted $\mathbb{Y}_{\kappa n}$.
If the system was isolated, these quantities would be conserved.
But the system is in contact with a set $\mathcal{P}=\{\rho\}$ of equilibrium \emph{reservoirs}. 
Each reservoir $\rho$ is characterized by a number of intensive entropic fields~\cite{callen1985thermodynamics} 
conjugated with the extensive quantities exchanged with the system, e.g. inverse temperature with energy, chemical potential divided by temperature with particle number, etc.).
These fields are denoted by $f_{y}$ with $y=(\rho, \kappa)$ and the set of all entropic fields by $\mathcal{Y}=\{(\rho, \kappa)\}$.
The reservoirs can trigger a set $\mathcal{E}^{+}=\{e\}$ of (directed) transitions between the mesoscopic states of the system $n \xrightarrow{e} n'$, where $n,n'$ are two different states, and $e$ is the transition mechanism. 
Mesoscopic states and transition mechanisms define the nodes and edges of a directed graph $(\mathcal{V},\mathcal{E}^{+})$, which allows for multiple edges between pairs of nodes. Micro-reversibility requires that each oriented edge $e$ has a corresponding inverse transition denoted $-e$. Using the notation $\mathrm{o}(e)$ to select the origin node of the edge $e$, the components of the \emph{incidence matrix} that identifies the multigraph are $\mathbb{D}_{ne}=\delta_{n\mathrm{o}(-e)}-\delta_{n \mathrm{o}(e)}$.
%\begin{align}
%    \mathbb{D}_{ne}=\delta_{n\mathrm{o}(-e)}-\delta_{n \mathrm{o}(e)}\,.
%\end{align}
Since the extensive quantities are conserved in the union of system plus reservoirs, they lead to balance equations along each transition $e$ 
\begin{align}
   \mathbb{Y}_{\mathrm{o}(-e) \kappa} - \mathbb{Y}_{\mathrm{o}(e) \kappa} = \sum_{m} \mathbb{Y}_{\kappa m} \mathbb{D}_{me} = \sum_\rho \mathbb{X}_{(\rho, \kappa)e}\,,  \label{eq:extensive_balance_intro}
\end{align}
where the matrix $\mathbb{X}=(\mathbb{X}_{ye})$ encodes the amount of extensive quantities exchanged with each such reservoir $y$ along $e$.  Micro-reversibility is formalized as the requirement that $\mathbb{X}_{y,-e} \coloneq - \mathbb{X}_{ye}$.
In addition to the $\abs{\mathcal{K}}$ \emph{trivial} conservation laws leading to the balance equation \eqref{eq:extensive_balance_intro}, additional \emph{non-trivial} conservation laws may arise from constraints in the internal structure of the system.

The dynamics of the system obeys a continuous-time Markov process over the set of mesoscopic states.
This implies that the probability vector $\vec{p}(t)=(p_n(t))$ describing the probabilities to find the system in each mesoscopic state at time $t$ is the solution of the master equation generated by the $\abs{\mathcal{V}} \times \abs{\mathcal{V}}$ matrix of transition rates $\mathbb{W}(t)$:
\begin{align}
	\d_t\vec{p}(t) = \mathbb{W}(t) \vec{p}(t) 
    = \mathbb{D} \vec{J}(t) \, , \label{eq:me}
\end{align}
where the component of the vector of probability currents $\vec{J}(t)$ read
\begin{align}
  J_{e}(t) \coloneq w_{e}(t) p_{\mathrm{o}(e)}(t) - w_{-e}(t)p_{\mathrm{o}(-e)}(t) = - J_{-e}(t)\,. \label{eq:prob_curr_def}
\end{align}
%Note that the sign of the current along each transition mechanism depends on the choice of the orientation for $\mathcal{E}^{+}$. The definition~\eqref{eq:prob_curr_def} implies that the currents along the reversed transition paths are $J_{-e}= - J_{e}$. 
The explicit time dependence will often be omitted in the rest of the paper.
%
%\iffalse
Dynamically, the balance equations for the extensive quantities~\eqref{eq:extensive_balance_intro} give rise to the continuity equations
\begin{align}
    \d_t \avg{\mathbb{Y}_{\kappa}}\coloneq \d_t \sum_n \mathbb{Y}_{\kappa n} p_n = \avg{\d_t \mathbb{Y}_{\kappa}} 
    %+ \sum_{e \rho} \mathbb{X}_{(\rho, \kappa)e} J_{e} \;,
    + \sum_{\rho} \vec{I}_{(\rho, \kappa)} \;, \label{eq:balance_trivial}
\end{align}
where the last term represents the currents of extensive quantities exchanged with the reservoirs
\begin{align}
    \vec{I}=\mathbb{X}\vec{J}\,.
    \label{eq:physcurrentdef}
\end{align}

The thermodynamic consistency is ensured by the \emph{local detailed balance} (LDB) which links the stochastic dynamics and thermodynamics (the Boltzmann constant $k_\mathrm{B}=1$) \cite{bergmann1955new, esposito2012stochastic, maes2020local, falasco2021local}:
\begin{align}
\ln\frac{w_{e} }{w_{-e} } = (\mathbb{D}\tr \vec{S} -\mathbb{X}\tr \vec{f})_{e} \,. \label{eq:ldb}
\end{align}
Here $\vec{S}$ denotes the internal entropy of the mesoscopic states. 
As a result, the first term in the r.h.s. of~\eqref{eq:ldb} denotes the internal entropy change arising in the system along the transition $e$, while the second one denotes the entropy changes in the reservoirs caused by the exchanges of extensive quantities.

By rewriting the probability currents~\eqref{eq:prob_curr_def} as
\begin{align}
J_{e} =w_{e} p_{\mathrm{o}(e)} \left( 1-\e^{-A_{e}}\right)\,, \label{eq:curr_vs_affinity}
\end{align}
we introduce the edge affinities
\begin{align}
	A_{e} \coloneq \ln \frac{w_{e} p_{\mathrm{o}(e)}}{w_{-e}p_{\mathrm{o}(-e)}}  \label{eq:aff}
\end{align}
which, using the LDB condition~\eqref{eq:ldb}, take the form
\begin{align}
  \vec{A} = -\mathbb{X}\tr \vec{f} + \mathbb{D}\tr(\vec{S} - \ln \vec{p}) \, . \label{eq:affities_fields}
\end{align}
Now and in the following, component-wise operations over vectors will be used: $\vec{a}\vec{b}~=~(a_i b_i)$,  $\ln \vec{a}=(\ln a_i)$ and $\vec{a}\vec{b}\tr=(a_ib_j)$.
The affinities~\eqref{eq:affities_fields} represent the entropy changes in the reservoirs and in the system (internal and self-information) induced by the transition $e$.
The EPR along each transition can therefore be expressed as the flux--force expression
\begin{align}
 \dot{\vec{\sigma}} &= \vec{J} \vec{A} \geq 0 \, , \label{eq:edge_epr}
\end{align}
and the total EPR by  
\begin{align}
\dot\Sigma (t) = \sum_{e} \dot{\sigma}_{e} (t) =  \sum_{e} J_{e}(t) A_{e}(t)
    = \vec{J}\tr \vec{A} \geq 0\,. \label{eq:epr}
\end{align}
Indeed, using~\eqref{eq:affities_fields} and~\eqref{eq:ldb}, and introducing the average entropy of the system $S_\mathrm{sys} \coloneq \sum_n (S_n - \ln p_n) p_n$, we find as expected \cite{esposito2012stochastic} that
\begin{align}
    \d_t S_\mathrm{sys} 
    = \dot\Sigma +
    \underbrace{\sum_n p_n d_t S_n - \sum_{e,y} f_y \mathbb{X}_{ye} J_{e}}_{\text{Entropy flow }} \,.
\end{align}
When the system is at equilibrium, the affinities, the currents and the EPR along every edge $e$ vanish, which coincides with the condition of detailed balance  
\begin{align}
A_{e} =0 \iff w_{e} p_{\mathrm{o}(e)}^{\mathrm{eq}} = w_{-e}p_{\mathrm{o}(-e)}^{\mathrm{eq}} \iff J_{e}=0 \iff \dot{\vec{\sigma}}=0 \label{eq:db}\,.
\end{align}

Master equations of the type \eqref{eq:me} with \eqref{eq:ldb} are widely used, for instance to model chemical and biological systems~\cite{altaner2015, rao2018chemical, forastiere2020strong} or electric circuits~\cite{esposito2007entropy, pekola2015towards,  freitas2021stochasticII}.

\subsection{Topology and conservation laws}
\label{sec:conservationlaws}

Introducing the incidence matrix $\mathbb{D}$ in Eq.~\eqref{eq:me} enables the deployment of topological tools. Indeed, Schnakenberg proved that cycles of transitions form a basis for the space of the steady-state probability currents~\cite{schnakenberg1976network}, and in Ref.~\cite{rao2018conservation} these cycles were used to identify all the conservation laws at work in the system. We briefly review these results.

\begin{figure}[t]
    \subfloat[][Scheme]
    { \def\svgwidth{0.31\textwidth}
      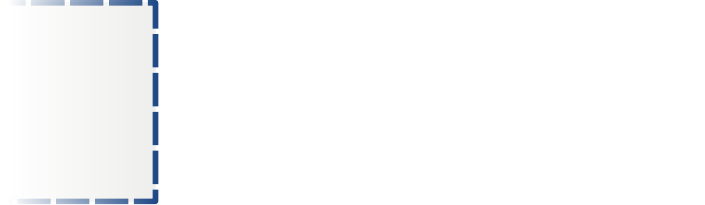
    }
    \subfloat[][Network]
    { \def\svgwidth{0.12\textwidth}
      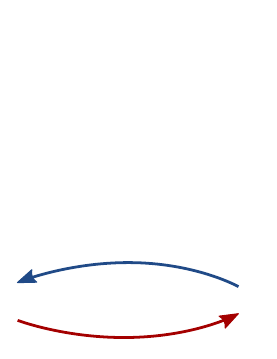
    }
    \caption{        The single QD connected with two reservoirs
      (a) Schematic representation
      (b) Network representation: the system can switch between the states $0,1$ through the transitions $+1 $ and $+2$. }
    \label{fig:transport}
\end{figure}

%We will now see how cycles contribute to the thermodynamics of non-stationary systems exchanging physical extensive quantities with an reservoirs (thus giving rise to physical currents between reservoirs). 
The cycle decomposition of the Markov process is obtained by finding the kernel of the incidence matrix $\mathbb{D}$, \textit{i.e.} the space of vectors satisfying $\mathbb{D}\vec{C}_{\alpha}=\vec{0}$. Physically, these vectors identify cycles since they are sequences of transitions such that after completion one returns back to the initial state. The kernel $\mathcal{A} \equiv \ker \mathbb{D}$ is spanned by $|\mathcal{A}|$ vectors $\vec{C}_\alpha$ which are the column vectors of the matrix $\mathbb{C}=(\mathbb{C}_{e \alpha})$. They are  such that $\mathbb{C}_{e\alpha}=0$ if the cycle $\alpha$ does not contain the edge $e$ and $\mathbb{C}_{e\alpha} = \pm 1$ otherwise. The sign defines the orientation needed to complete the cycle (see the networks represented in Figs. \ref{fig:transport} or \ref{fig:solarcell}). 

%While these cycles are useful to obtain the steady-state distribution of a Markov process~\cite{schnakenberg1976network}, they are not fundamental objects for thermodynamics. By this we mean that if they are connected to the same reservoirs, the exchanged extensive quantities may be the same for different cycles $\vec{C}_{\alpha}$~\cite{polettini2016conservation, rao2018conservation}, and thus this description may become redundant.

%The redundancy is eliminated if we use a representation of the topological properties of the network that also takes into account the extensive quantities exchanged along each cycle. This representation is obtained constructing the matrix

While cycles are enough to obtain the steady-state distribution of a Markov process~\cite{schnakenberg1976network}, the extensive quantities exchanged along each cycle are of relevance for thermodynamics. They are encoded in the \emph{physical topology} matrix
\begin{align}
    \mathbb{M}\coloneq \mathbb{X} \mathbb{C}\,. \label{eq:physical_topology_M}
\end{align}
Each entry represents the amount of extensive quantity $\kappa$ exchanged with physical reservoir $\rho$ after performing a cycle $\alpha$. The constraints that the conservation laws impose on the cycle affinities are encoded in the left null space $\Lambda$. This vector space is spanned by $|\Lambda|=\abs{\mathrm{coker}\, \mathbb{M}}$ conservation law vectors  $\vec{\ell}_{\lambda}$ which are the columns vectors of the matrix $\mathbb{l}\coloneq(\mathbb{l}_{y \lambda})$ that satisfies
\begin{align}
\mathbb{l}\tr \mathbb{M}=\mathbb{0}\,. \label{eq:cl_definition}
\end{align}

For each conservation law $\vec{\ell}_\lambda$ we can identify a corresponding conserved quantity $\vec{L}_\lambda$, column vector of the matrix $\mathbb{L}=(\mathbb{L}_{n\lambda})$, that satisfies the balance equation
\begin{align}
  \mathbb{D}\tr \vec{L}_\lambda =  \mathbb{X}\tr\vec{\ell}_{\lambda} \, ,\label{eq:conservationlaws}
\end{align}
or equivalently in matrix form
\begin{align}
  \mathbb{D}\tr \mathbb{L} = \mathbb{X}\tr\mathbb{l} \, . \label{eq:conservationlaws_matrix}
\end{align}
This result is obtained realizing that the definition \eqref{eq:cl_definition} is equivalent to finding the subspace orthogonal to $\ker \mathbb{D}=\mathrm{span}\, \{\vec{C}_{\alpha}\}$ because of \eqref{eq:physical_topology_M}.
This means that the vectors $\vec{\ell}_{\lambda}\tr \mathbb{X}$ are in the cokernel of the incidence matrix $\mathbb{D}$, \textit{i.e.}  $\vec{\ell}_{\lambda}\tr \mathbb{X} \in \mathrm{coker}\,\mathbb{D}=(\ker \mathbb{D})^{\perp}$.
It follows that, since $(\ker \mathbb{D})^{\perp}$ is isomorphic to the image of $\mathbb{D}\tr$ (the coimage of $\mathbb{D}$, see \emph{e.g.} \cite{horn2012matrix}), there exist $\abs{\Lambda}$ basis vectors $\vec{L}_{\lambda}$ that are mapped into the transposed of $\vec{\ell}_{\lambda}\tr \mathbb{X}$ by $\mathbb{D}\tr$.
Physically,  Eqs.~\eqref{eq:conservationlaws} and \eqref{eq:conservationlaws_matrix} are balance equations.
The l.h.s. quantify the changes of the conserved quantities in the system when transitions occur.
The r.h.s. identifies the contributions to these variations due to the different reservoirs. Notice that each $\vec{L}_\lambda$ is defined up to an additive constant, as only their differences enter the balance equation. 
Importantly, these conserved quantities $\mathbb{L}$ encompass the $\abs{\mathcal{K}}$ \emph{trivial} conservation laws, but also contain $\abs{\Lambda} - \abs{\mathcal{K}} \ge 0$ \emph{non-trivial} additional ones, which are system specific (see Examples \ref{sec:ex_2} and \ref{sec:ex_3}).

\subsection{Fundamental forces}

The balance equation \eqref{eq:conservationlaws} can be used to split the vector of entropic fields $\vec{f}$ into one block of potential fields $\vec{f}_{\mathrm{p}}$ and one of fundamental nonconservative fields $\vec{f}_\mathrm{f}$, $\vec{f} = \left(\vec{f}_\mathrm{f}, \vec{f}_\mathrm{p}\right)\tr$. Physically, this corresponds to selecting a minimal subset $\vec{f}_{\mathrm{p}}$ of entropic fields that defines a reference equilibrium for the system, and the subset of remaining fields $\vec{f}_{\mathrm{f}}$, whose presence may independently prevent the system from reaching such equilibrium.
The potential fields $\vec{f}_{\mathrm{p}}$ are obtained by selecting the maximal invertible submatrix of $\mathbb{l}$, denoted $\mathbb{l}_{\mathrm{p}}$. This matrix is always square, has dimension $|\Lambda|$, and is full rank, since from the definition  Eq.~\eqref{eq:cl_definition} it follows that $\mathrm{rk}\, \mathbb{l}=|\Lambda|$. We call $\mathbb{l}_{\mathrm{f}}$ the remaining rectangular block of the matrix $\mathbb{l}$. 
Eq.~\eqref{eq:conservationlaws_matrix} can be solved for the exchanged quantities corresponding to $\vec{f}_{\mathrm{p}}$, denoted $\mathbb{X}_{\mathrm{p}}$, in terms of $\mathbb{l}_{\mathrm{p}}$, $\mathbb{l}_{\mathrm{f}}$, and the remaining $\abs{\mathcal{Y}} - \abs{\Lambda}$ exchanged quantities, $\mathbb{X}_{\mathrm{f}}$, with $\mathbb{X} = \left(\mathbb{X}_{\mathrm{f}}, \mathbb{X}_{\mathrm{p}}\right)$:
\begin{align}
\mathbb{X}_{\mathrm{p}}\tr =  - \mathbb{X}_{\mathrm{f}}\tr \mathbb{l}_{\mathrm{f}} \, \mathbb{l}_{\mathrm{p}}^{-1} + \mathbb{D}\tr\mathbb{L}\mathbb{l}_{\mathrm{p}}^{-1}\,. \label{eq:expl_}
\end{align}
This procedure leads to the following rewriting of the exchange contribution appearing in the LDB~\eqref{eq:ldb}
\begin{align}
  \mathbb{X}\tr \vec{f}= - \mathbb{X}_{\mathrm{f}}\tr \vec{\mathcal{F}} + \mathbb{D}\tr \mathbb{L}  (\mathbb{l}^{-1}_{\mathrm{p}}) \vec{f}_{\mathrm{p}}\,. \label{eq:splitting_flat}
\end{align}
Eq.~\eqref{eq:splitting_flat} features the \emph{nonconservative forces} defined by
\begin{align}
\vec{\mathcal{F}} \coloneq - \vec{f}_{\mathrm{f}} + \mathbb{l}_{\mathrm{f}} \, \mathbb{l}^{-1}_{\mathrm{p}} \vec{f}_{\mathrm{p}}\,, \label{eq:fundForce}
\end{align}
which we refer to as \emph{fundamental} because they are a minimal set of independent mechanisms that break detailed balance, \emph{i.e.} prevent the system from reaching equilibrium. For example, in a system with two reservoirs exchanging energy, the vector $\vec{\mathcal{F}}$ contains the difference between the inverse temperatures of the reservoirs (see also Example~\ref{sec:ex_2}).

Note that, because of the rank-nullity theorem for the $\abs{\mathcal{Y}}\times\abs{\mathcal{A}}$ physical topology matrix $\mathbb{M}$, we have that
\begin{align}
\abs{\text{Fund. forces}} = \abs{\mathcal{Y}} - \abs{\Lambda} = \mathrm{rk}\, \mathbb{M} = \abs{\mathcal{A}} - \abs{\ker \mathbb{M}} \, , \label{eq:rk_nullity}
\end{align}
i.e. the number of fundamental forces is given by the rank of $\mathbb{M}$.
When $\mathrm{rk}\, \mathbb{M} = 0$: $\abs{\mathcal{Y}} = \abs{\Lambda}$; there are not enough reservoirs for nonconservative forces to be generated; and the system is unconditionally detailed balance.
This further emphasizes how $\mathbb{M}$ encapsulates the thermodynamic properties of the system.

Finally,  we make explicit that Eq.~\eqref{eq:fundForce} defines a linear transformation from the entropic intensive fields $\vec{f}$ to the fundamental forces and potential fields:
\begin{align}
    \mathbb{T} \vec{f}=
  \begin{pmatrix}
    \vec{\mathcal{F}}\\
    \vec{f}_{\mathrm{p}}
    \end{pmatrix}
	\, ,
	\label{eq:decomposition}
\end{align}
with its block form being
\begin{align}
  \mathbb{T} =
  \begin{pmatrix}
    -\mathbb{1} & \mathbb{l}_{\mathrm{f}} \, \mathbb{l}^{-1}_{\mathrm{p}}\\
    \mathbb{0} & \mathbb{1}\\
\end{pmatrix} \, . \label{eq:splitting}
\end{align}
Notice that the transformation matrix \(\mathbb{T}\) is involutory, \emph{i.e.} $\mathbb{T}^2=\mathbb{1}$.

\subsection{EPR and thermodynamic potentials}

The interplay between topological and thermodynamic properties of the system also shapes the expression of the EPR.
Indeed, using Eq.~\eqref{eq:splitting_flat}, the edge affinities~\eqref{eq:affities_fields} and the entropy production~\eqref{eq:edge_epr} can be rewritten as  
\begin{align}
  \vec{A} =  \mathbb{X}_{\mathrm{f}}\tr \vec{\mathcal{F}}  + \mathbb{D}\tr \vec{\Phi} \,, \label{eq:affities_fields_Fund}
\end{align}
and
\begin{align}
 \dot{\vec{\sigma}} &= \vec{J} \vec{A} = \vec{J}  \left(\mathbb{X}_{\mathrm{f}}\tr \vec{\mathcal{F}}  + \mathbb{D}\tr \vec{\Phi}\right)\,, \label{eq:edge_epr_cl}
\end{align}
respectively, where the stochastic Massieu potential
\begin{align}
	\vec{\Phi} &\coloneq \vec{\phi} - \ln \vec{p} \label{eq:distribution-general}
\end{align}
is obtained by complementing the Massieu potential of the states
\begin{align}
  \vec{\phi}
  &\coloneq
  \vec{S} - \mathbb{L}(\mathbb{l}^{-1}_{\mathrm{p}}) \vec{f}^{\vphantom{-1}}_{\mathrm{p}}  \label{eq:potential}
\end{align}
with the self-information $-\ln \vec{p}$.
The first contribution in Eqs.~\eqref{eq:affities_fields_Fund} and \eqref{eq:edge_epr_cl} describes the dissipative contribution of the nonconservative forces, whereas the second that of the potential ones.

Using Eq.~\eqref{eq:edge_epr_cl}, the full EPR~\eqref{eq:epr} can be thus rewritten as
\begin{align}
\dot{\Sigma} = \d_t \Phi + \vec{\mathcal{F}}\tr \vec{I}_{\mathrm{f}} + \dot{\nu} \, , \label{eq:eprdecomposition}
\end{align}
where
\begin{align}
  \Phi \coloneq  \vec{\Phi}\tr \vec{p}
\end{align}
is a \emph{nonequilibrium Massieu potential} for the system, 
\begin{align}
  \vec{I}_{\mathrm{f}} \coloneq \mathbb{X}_{\mathrm{f}} \vec{J}
  \label{eq:physcurrents}
\end{align}
denotes the current of system quantities conjugated to the fundamental forces, and the residual term
\begin{align}
    \dot{\nu}&\coloneq - \left(\d_t \vec{\phi}\right)\tr\vec{p}
\end{align}
denotes the average of the variation in time of the Massieu potential of the states.
Equation~\eqref{eq:eprdecomposition} discriminates the three core mechanisms that contribute to dissipation.
The first term describes the dissipation associated to transient relaxation effects, and it vanishes at steady state.
The second term describes the dissipation due to nonconservative flows of system quantities through the system, and it is the only nonvanishing term for a NESS.
Finally, the third term in Eq.~\eqref{eq:eprdecomposition} is a \emph{driving contribution} which describes the dissipation due to external time-dependent drivings.

In detailed-balanced systems---\textit{i.e.} vanishing forces, \(\vec{\mathcal{F}} = 0\)---with no time-dependent driving, the EPR is fully characterized by the changes of nonequilibrium Massieu potential, $\Phi$.
Under these conditions, this potential becomes a Lyapunov function which keeps increasing as the system relaxes to equilibrium.
To review this fact, let us first note that the state
\begin{align}
	\vec{p}^\mathrm{eq} = \left(p_n^\mathrm{eq}\right) = \left( \exp\{\phi_n - \Phi_\mathrm{eq}\} \right) \, ,
    \label{eq:gibbs}
\end{align}
with $\Phi_\mathrm{eq} = \ln \sum_n \exp\{\phi_n\}$ is the equilibrium state of the system, since it satisfies the detailed balance property \eqref{eq:db} (consider Eq.~\eqref{eq:affities_fields_Fund} with \(\vec{\mathcal{F}} = 0\)).
Using this last expression, we obtain
\begin{equation}
	\Phi = \left( \vec\phi - \ln \vec{p} \right)\tr \vec{p} = \Phi_{\mathrm{eq}} - D_{\mathrm{KL}}\left( \vec{p} \| \vec{p}^{\mathrm{eq}} \right) \, ,
	\label{eq:massieu_dkl_relation}
\end{equation}
where we introduced the Kullback--Leibler divergence $D_{\mathrm{KL}}\left( \vec{p} \| \vec{p}^{\mathrm{eq}} \right) = \vec{p}\tr \ln \left(\frac{ \vec{p}}{\vec{p}^\mathrm{eq}}\right) \ge 0$, which is non-negative and vanishes solely at equilibrium.
Using the fact that \(\Phi_{\mathrm{eq}}\) is constant, \( \d_{t} D_{\mathrm{KL}} = - \dot\Sigma \le 0 \), we recover the role of \(\Phi\) as a Lyapunov function.

We conclude this section by writing the continuity equation constraining the evolution in time of the average of each conserved quantity,
\begin{align}
    \d_t \avg{L_{\lambda}} = \left(\d_t \vec{L}_\lambda\right)\tr\vec{p} + \vec{L}\tr_\lambda  \d_t \vec{p} = \left(\d_t \vec{L}_\lambda\right)\tr\vec{p} + \vec{\ell}_\lambda\tr \vec{I} \, , \label{eq:balance_eq_currents}
\end{align}
which is obtained using the master equation \eqref{eq:me} together with Eqs.~\eqref{eq:physcurrentdef},~\eqref{eq:conservationlaws}.
This equation clarifies how the changes of \(\avg{L_{\lambda}}\) can be due to either intrinsic variations of \(\vec{L}_\lambda\) due to time-dependent driving (the first term on the r.h.s.) or to exchanges with the reservoirs (the second term on the r.h.s.). Eq.~\eqref{eq:balance_eq_currents} generalizes Eq. \eqref{eq:balance_trivial} to the case of non-trivial, system-dependent conserved quantities.

% =============================================================
\section{Linear Regime with Protocols acting on the Reservoirs}
\label{sec:driving_reservoirs}

In this Section we consider protocols that only act on the entropic fields characterizing the state of the reservoirs, $\vec{f}$, and do not affect the matrix $\mathbb{X}$ of physical quantities exchanged with the reservoirs. 
We first obtain the linearized solution of the master equation \eqref{eq:me} in terms of the fundamental forces and Massieu potentials, \S\ref{sec:linearized_dynamics}.
In \S\ref{sec:detailed_balance} we consider detailed-balanced dynamics to show how Green--Kubo relations and thermodynamic stability conditions are recovered from our approach.
In \S\ref{sec:quadraticepr_fields} we consider generic periodic drivings and show that the EPR in steady conditions can be written as the modulus of the protocol amplitudes with respect to a suitable scalar product.
The matrix representing this scalar product is used to obtain a generalized Onsager matrix in \S\ref{sec:generalized_onsager_fields}.
This is the main result of this Section.
Finally, in \S\ref{sec:currents_to_intensive} and \S\ref{sec:relationResponseMatrices} we analyze the response of the currents and the balance equations under this type of protocols.

\subsection{Dynamical Response of Generic Systems}
\label{sec:linearized_dynamics}

To characterize the dynamical response of systems close to equilibrium, we first establish how the currents become linear in the edge affinities.
Let the fields \(\vec{f}^{\mathrm{eq}}\) identify some reference detailed balanced conditions described by transition rates \(\vec{w}^{\mathrm{eq}}\) and whose equilibrium probability distribution is \(\vec{p}^{\mathrm{eq}}\).
Upon an instantaneous and small displacement of these fields, $\vec{f}(t)=\vec{f}^{\mathrm{eq}} + \delta \vec{f}(t)$, the transition rates are displaced as $\vec{w}(t) \approx \vec{w}^{\mathrm{eq}} + \delta \vec{w}(t)$, and in turn the probability responds as $\vec{p}(t) \approx \vec{p}^{\mathrm{eq}} + \delta \vec{p}(t)$, where in $\delta\vec{p}$ and $\delta\vec{w}$ only contributions linear in $\delta \vec{f}$ have been retained. No probability current flows at equilibrium. 
Denoting by \(\vec{j}^{\mathrm{eq}} \coloneq \vec{w}^{\textrm{eq}} \vec{p}^{\textrm{eq}}\) the equilibrium fluxes, the first order contribution of the current \eqref{eq:curr_vs_affinity} can be written as
\begin{align}
	\vec{J} \approx \delta \vec{J} = \vec{j}^{\mathrm{eq}} \delta \vec{A} \, ,
	\label{eq:linearizedcurrent}
\end{align}
where the linearized affinities are obtained from Eq.~\eqref{eq:affities_fields}, exploiting the identity for the linear  corrections $\delta \ln \vec{p}=\delta \vec{p}/\vec{p}^{\mathrm{eq}}$:
\begin{align}
	\vec{A} \approx \delta \vec{A} = - \mathbb{X}\tr \delta \vec{f} - \mathbb{D}\tr \delta \ln \vec{p} \, .
	\label{eq:linearizedaffinity_result}
\end{align}
Equivalently, using Eq.~\eqref{eq:aff} the affinities can be expressed in terms of the response of the rates
\begin{align}
	\delta \vec{A} &= \left(
		\frac{\delta w_{e}}{w^{\textrm{eq}}_{e}}
		- \frac{\delta w_{-e}}{w^{\textrm{eq}}_{-e}}
		- \sum_n \mathbb{D}_{e,n} \frac{\delta p_n}{p_n^{\textrm{eq}}}
	\right) \, ,
	\label{eq:linearizedaffinity}
\end{align}
which in turn respond as
\begin{align}
	\delta w_{e} &= \sum_y \pder[w_{e}]{f_y} \delta f_y=\sum_y \left(\pder[\ln s_e]{f_y} - \frac{1}{2}
	%\left( \mathbb{X} - \mathbb{D}\tr \frac{\partial\,\vec{S}}{\partial \vec{f}}\right)_{ey}
	\mathbb{X}_{ey}
	\right) w_{e}^{\mathrm{eq}} \delta f_y \, .
	\label{eq:rate_perturbed}
\end{align}
For this last expression, we used the LDB~\eqref{eq:ldb} to write the rates in the form
\begin{align}
	w_{e} = s_{e} \exp\left\{\frac{1}{2}\left( -\mathbb{X}\tr \vec{f} +\mathbb{D}\tr \vec{S}\right)_{e}\right\}\,,
	\label{eq:rates}
\end{align}
with $s_{e}\coloneq\sqrt{w_{e} w_{-e}}$ being a symmetric prefactor invariant under the exchange of $e$ with $-e$.
Without loss of generality, we only consider protocols such that $\delta \vec{S} = \sum_y \pder[\vec{S}]{f_y} \delta f_y = 0$, as the additional term could be treated as a perturbation in exchanged quantities $\mathbb{X}$, considered in Sec.~\ref{sec:generalprotocol}.

Equations~\eqref{eq:linearizedcurrent}, \eqref{eq:linearizedaffinity_result} and \eqref{eq:linearizedaffinity} establish the linear response of currents and affinities.
Note however that the response is written in terms of two physically distinct contributions:
The first one is due to the variation of the external intensive fields, $\delta \vec{f}$, while the second is a dynamic response and only depends on the current state of the system, which is represented by the instantaneous relative deviation from the equilibrium distribution, $\delta \ln \vec{p} = \delta \vec{p}/\vec{p}^\mathrm{eq}$.

Before we proceed with our derivation of the linearized master equation, we introduce the Hermitian scalar product $\avg{\cdot \,, \cdot}$ over the space of edge vectors defined  (in matrix notation and by components resp.) by
\begin{align}
\avg{\vec{a}, \vec{b}} \coloneq \vec{a}^\dagger (\vec{j}^{\mathrm{eq}}  \vec{b}) = \sum_{e}  j^{\mathrm{eq}}_{e} a_{e}^* b_{e}\, , \label{eq:hermitian_product}
\end{align}
which corresponds to an equilibrium average (with equilibrium fluxes as weights over the edges instead of probabilities).  With a slight abuse of notation, we will use the same symbol for the matrix obtained from the weighted contraction of two matrices over edges defined by
\begin{align}
\avg{\mathbb{M},\mathbb{N}}_{ij} \coloneq  \sum_{e} j^{\mathrm{eq}}_{e} \mathbb{M}^*_{i,e}  \mathbb{N}_{e,j} \,. \label{eq:hermitian_product_matrix}
\end{align}
It will be clear from the context if the result of the operation is a scalar or a matrix.
Finally note that the above definitions imply $\avg{\mathbb{M}\vec{a}, \mathbb{N}\vec{b}}= (\mathbb{M}\vec{a})^\dagger (\vec{j}^{\mathrm{eq}}\mathbb{N}\vec{b})=\vec{a}^{\dagger} \avg{\mathbb{M}, \mathbb{N}} \vec{b}$. This definition bears connections to the one introduced in Ref. \cite{polettini2013nonconvexity}.

The linearized version of master equation~\eqref{eq:me}, can thus be written as
\begin{align}
	\d_t \delta \vec{p}&=  \mathbb{D} \, \delta \vec{J} = \avg{\mathbb{D}\tr, \delta \vec{A}} \label{eq:linearizedme_affinity}\\
 &= -\avg{\mathbb{D}\tr, \mathbb{X}\tr} \delta \vec{f} - \avg{\mathbb{D}\tr,\mathbb{D}\tr} \delta \ln \vec{p}\label{eq:linearizedme}
\end{align} 
where we used Eqs.~\eqref{eq:linearizedcurrent} and \eqref{eq:linearizedaffinity_result}.
In addition, using Eq.~\eqref{eq:affities_fields_Fund}, we can establish the alternative formulation in terms of the nonconservative forces and the Massieu potential,
\begin{equation}
    \d_t \delta \vec{p} = \avg{\mathbb{D}\tr, \mathbb{X}_\mathrm{f}\tr} \delta \vec{\mathcal{F}} + \avg{\mathbb{D}\tr,\mathbb{D}\tr} \left( \delta \vec{\phi} - \delta \ln \vec{p} \right) \label{eq:linearizedme_decomposed} \, ,
\end{equation}
which follows from rewriting the linear contribution to the affinities \eqref{eq:linearizedaffinity_result} as
\begin{align}
	\delta \vec{A} = \mathbb{X}_{\mathrm{f}}\tr \delta\vec{\mathcal{F}} + \mathbb{D}\tr \left( \delta \vec{\phi} - \delta \ln \vec{p}\right) \, ,
	\label{eq:generalaffinities}
\end{align}
with
\begin{align}
	\delta \vec{\mathcal{F}} & = - \delta \vec{f}_{\mathrm{f}} + \mathbb{l}_{\mathrm{f}} \, \mathbb{l}^{-1}_{\mathrm{p}} \delta \vec{f}_{\mathrm{p}} \\
	\delta \vec{\phi} & = - \mathbb{L}(\mathbb{l}^{-1}_{\mathrm{p}}) \delta \vec{f}^{\vphantom{-1}}_{\mathrm{p}}
	\, . \label{eq:deltaphi:fieldsPert}
\end{align}
We remark that the perturbations acting on the symmetric part of the rates $s_e$ do not contribute to $\delta \vec{p}$ near equilibrium. In other words, a perturbation on the symmetric part of the rates does not influence the linear response of the system \cite{ maes2010rigorous, baiesi2013response}.

The solution of the linearized master equation \eqref{eq:linearizedme} for a general perturbation over the intensive fields $\vec{f}$ provides an ensemble description for near-equilibrium mesoscopic systems~\cite{mclennan1959statistical, maes2010rigorous} even when detailed balance is violated.
This description is readily obtained in Fourier transform, which we recall here for convenience
\begin{align}
\hat{g}(\omega)\coloneq \frac{1}{2 \pi} \int_{-\infty}^\infty \d \tau \, \e^{-\i \omega \tau}\check{g}(\tau) \, , \label{eq:fourier}
\end{align}
(the corresponding inverse transform being $\check{g}(t)= \int_{-\infty}^{\infty} \d \omega \e^{\i \omega t} \hat{g}(\omega)$).
Indeed, by solving for the Fourier transform of the instantaneous deviations $\delta \ln \hat{\vec{p}}(\omega)$ in terms of the perturbation $\delta \hat{\vec{f}}(\omega)$, we obtain
\begin{align}
  \delta \ln \hat{\vec{p}}(\omega)   &= \frac{\delta \hat{\vec{p}}(\omega)}{\vec{p}^\mathrm{eq}} =
  -\mathbb{A}(\omega) \avg{\mathbb{D}\tr, \mathbb{X}\tr}\delta \hat{\vec{f}}(\omega) \, , \label{eq:fourier_me_solution}
\end{align}
where
\begin{align}
    \mathbb{A}(\omega) \coloneq \left(\i \omega \mathbb{P} + \avg{\mathbb{D}\tr, \mathbb{D}\tr}\right)^{-1} \, ,
    \label{eq:auxiliarymatrA}
\end{align}
and $\mathbb{P}_{nm} \coloneq \delta_{nm} p^{\mathrm{eq}}_m$.
The auxiliary matrix \(\mathbb{A}(\omega)\) encodes how the probability vector responds to different frequencies.
We can thus characterize the solution in time domain
\begin{align}
  \vec{p}(t)   &\approx \vec{p}^{\mathrm{eq}}\left( 1 -\int_{-\infty}^{\infty} \d \omega  \e^{\i \omega t} \mathbb{A}(\omega) \avg{\mathbb{D}\tr, \mathbb{X}\tr}\delta \hat{\vec{f}}(\omega)  \right) \label{eq:fourier_mclennan}\\
  &=\vec{p}^{\mathrm{eq}}\left( 1 +\int_{-\infty}^{\infty} \d \omega  \e^{\i \omega t} \mathbb{A}(\omega) 
  \left(\avg{\mathbb{D}\tr, \mathbb{X}_\mathrm{f}\tr} \delta \hat{\vec{\mathcal{F}}}(\omega) + \avg{\mathbb{D}\tr,\mathbb{D}\tr} \delta \hat{\vec{\phi}}(\omega) \right) \right)\,.
 \label{eq:fourier_mclennan_fund}
\end{align}
where we also used Eq.~\eqref{eq:affities_fields_Fund}.
This near-equilibrium solution of the master equation characterizes the dynamical response of the system to protocols with arbitrary time-dependence.

\subsection{Response of Unconditionally Detailed-balance Systems}
\label{sec:detailed_balance}

For unconditionally detailed-balanced systems, we can obtain simple expressions for both the response of the system quantities and the total dissipation during the relaxation to equilibrium.
These results will be later generalized to systems in which detailed balance is broken by nonconservative forces.

\subsubsection{Linear response theory for system quantities}

Let us consider an unconditionally detailed-balanced system that is initially prepared in an equilibrium steady state, \(\vec{p}^{\mathrm{eq}'}\), defined according to Eq.~\eqref{eq:gibbs}.
Without loss of generality, we can write $\phi_n = \phi_n^\mathrm{eq'} = S_n - L_{n\lambda} \, f'_\lambda$, as a one-to-one mapping can be constructed between $\mathcal{Y}$ and $\Lambda$ in absence of nonconservative forces---in other words, we regard \(\mathbb{l}_{\mathrm{p}}\) as similar to an identity matrix.
A small perturbation of the intensive fields, $\vec{f}=\vec{f}' + \delta \vec{f}$, causes a change $\delta \vec{p}(t)$ in the probability vector.
Since $\vec{\mathcal{F}}=\vec{0}$, by solving Eq.~\eqref{eq:linearizedme_decomposed} we get
\begin{align}
  \delta \vec{p}(t) &= \int_0^t \d s \exp\left\{-\frac{\avg{\mathbb{D}\tr,\mathbb{D}\tr}}{\mathbb{P}} [t-s]\right\} \avg{\mathbb{D}\tr,\mathbb{D}\tr} \, \delta \vec{\phi}(s)  \,. \label{eq:prob_response_db_time}
\end{align}
The matrix $\avg{\mathbb{D}\tr,\mathbb{D}\tr}$ enjoys the following important property
\begin{align}
    \avg{\mathbb{D}\tr,\mathbb{D}\tr}\mathbb{P}^{-1} = \left(\sum_e \frac{j_e^{\mathrm{eq}}}{p_m^{\mathrm{eq}}} \mathbb{D}_{ne}\mathbb{D}_{me} \right)= \left(\sum_{e} w_{-e}^{\mathrm{eq}} \mathbb{D}_{ne}\delta_{m\mathrm{o}(e)} -\sum_{e} w_{e}^{\mathrm{eq}}\mathbb{D}_{ne}\delta_{m\mathrm{o}(-e)} \right)=\left(\sum_e \frac{j_e^{\mathrm{eq}}}{p_n^{\mathrm{eq}}} \mathbb{D}_{me} \mathbb{D}_{ne}\right)= \left(\avg{\mathbb{D}\tr,\mathbb{D}\tr}\mathbb{P}^{-1}\right)\tr\,,
\end{align}
which follows from the definition of the incidence matrix $\mathbb{D}$ and from the detailed-balance condition $j_e^{\mathrm{eq}}=j_{-e}^{\mathrm{eq}}$.
The above property also justifies the notation used in the matrix exponential in \eqref{eq:prob_response_db_time} to stress the symmetry of the matrix at the exponent.
Equation~\eqref{eq:prob_response_db_time} shows how the instantaneous response of the probability vector is the outcome of the propagation of the changes of the Massieu potential of the states \(\delta\vec\phi\).
The propagator, in turn, is essentially determined by topological and thermodynamic properties of the system through \(\avg{\mathbb{D}\tr,\mathbb{D}\tr}\mathbb{P}^{-1}\).

\iffalse
Linear response coefficients are defined for generic state observables can be expressed as linear combinations of the conservation laws  $\vec{L}_\lambda$, so it is sufficient to study the response of these latter quantities.
\fi
We now turn our attention to the linear response coefficients of the conserved quantities, $\vec{L}_\lambda$, since these provide a fundamental description of the state of the system.
After the perturbation, the deviations of the average $\vec{L}_\lambda$ from their equilibrium value are given by
\begin{align}
\delta\avg{L_\lambda(t)} = \vec{L}_\lambda\tr \delta \vec{p} = \int_{0}^{t} \d s \avg{\vec{L}(t-s) \frac{\avg{\mathbb{D}\tr,\mathbb{D}\tr}}{\mathbb{P}} \, \delta \vec{\phi}(s)}_\mathrm{eq}  \,,  \label{eq:gk_dynamical}
\end{align}
where we have introduced the evolved observable
\begin{align}
\vec{L}_\lambda(t)\coloneq \vec{L}_\lambda\tr  \exp\left\{-\frac{\avg{\mathbb{D}\tr,\mathbb{D}\tr}}{\mathbb{P}} t\right\}\,,\label{eq:observable_evoluted}
\end{align}
and the equilibrium average must be intended as a component-wise multiplication by $\vec{p}^{\mathrm{eq}}$.

Equation~\eqref{eq:gk_dynamical} is the Agarwal fluctuation--response relation for the  conserved quantities in an unconditionally detailed balanced systems \cite{kubo1957statistical, mori1958statistical, green1960comment, agarwal1972fluctuation, marconi2008fluctuation, pigolotti2021}.
Since we are considering perturbations of systems in thermal equilibrium, one can show that the response matrix is proportional to the derivative of the self-correlation matrix.
In fact, since these perturbations only act on the intensive fields $\vec{f}$, we can compute explicitly $\delta \vec{\phi} = - \sum_\lambda \vec{L}_\lambda \delta f_\lambda $, and from Eq.~\eqref{eq:prob_response_db_time} we obtain the Green--Kubo relation
\begin{align}
  \frac{\delta\avg{L_{\lambda}(t)}}{\delta f_{\lambda'}(s)}
  &= \pder{t}\avg{\vec{L}_{\lambda}(t-s) \vec{L}_{\lambda'}}_{\mathrm{eq}}   \label{eq:DB_OnsagerRec}\,.
\end{align}
This formula immediately implies Onsager reciprocity
\begin{align}
    \frac{\delta\avg{L_{\lambda}(t)}}{\delta f_{\lambda'}(s)} =\frac{\delta\avg{L_{\lambda'}(t)}}{\delta f_{\lambda}(s)} \,,
\end{align}
because the exponential matrix appearing in \eqref{eq:observable_evoluted} is symmetric.

We have here discussed the linear response regime of conserved quantities for the simplest class of dynamics---\textit{viz.} unconditionally detailed balance---and recovered the celebrated Onsager's reciprocity relation.
However, this proof cannot be used for system in which the nonconservative forces (such as externally imposed gradients) are present:
Proofs of Onsager reciprocity in nonequilibrium steady states indeed require different methods~\cite{kurchan1998fluctuation, lebowitz1999gallavotti, schnakenberg1976network, hill1982linear, andrieux2004fluctuation}.
In Sec.~\ref{sec:generalized_onsager_fields} we will give the construction of response functions expressed in terms of fundamental thermodynamic forces~\eqref{eq:fundForce}.
Our approach will \emph{(i)} provide a natural connection with results from macroscopic thermodynamics, and \emph{(ii)} generalize the proof of Onsager symmetry to time-dependent nonconservative forces.

\subsubsection{Relaxation between different equilibrium states and thermodynamic stability}
\label{sec:relaxation}

We now specialize our discussion to purely relaxation processes and investigate the corresponding entropy dissipation.
In these procesess, the system transitions from one equilibrium state to another as the outcome of an instantaneous switch of the entropic fields from $\vec{f}'$ to $\vec{f} = \vec{f}' + \delta \vec{f}$ at $t=0$.
Since no nonconservative force is present---\emph{i.e.} $\vec{\mathcal{F}}=0$---the two contributions to the entropy production (Eq.~\eqref{eq:eprdecomposition}) are the difference of Massieu potential between the initial and final state and the work done by the driving mechanism to produce the switch.
The latter contribution is readily evaluated:
$\dot{\nu} = - (\d_t \vec{\phi})\tr \, \vec{p} = - \delta_{\mathrm{D}}(t) \sum_\lambda \delta f_\lambda \vec{L}_{\lambda}\tr \vec{p}^{\mathrm{eq}'}$, where \(\delta_{\mathrm{D}}(t)\) is a Dirac delta distribution.
%and the average is performed over the old equilibrium distribution---since, at $t=0$, the system has not had time to evolve to a different state yet.
Upon integration of the EPR \eqref{eq:eprdecomposition} with the help of \eqref{eq:massieu_dkl_relation} we obtain
\begin{align}
	\Sigma & \coloneq \int_{0}^\infty \d t \, \dot{\Sigma}(t) = \Phi_{\mathrm{eq}} - \Phi_{\mathrm{eq}'} - \sum_\lambda \delta f_\lambda L_{\lambda n}  p_n^{\mathrm{eq}'}
		= D_{\mathrm{KL}} \big( p^{\mathrm{eq}'} \big\| p^\mathrm{eq} \big) \, .
	\label{eq:epdkl}
\end{align}
The last equality follows from the definition \eqref{eq:distribution-general}, the fact that equilibrium Massieu potentials are constants. and the constraints on the conservation of probability, $\sum_n \delta p_n = 0$:
\begin{equation}
    \begin{aligned}
	\Phi_{\mathrm{eq}} & =
	\left( \vec{\phi} - \ln \vec{p}^{\mathrm{eq}} \right)\tr (\vec{p}^{\mathrm{eq}'} +\delta \vec{p}) =
	\left( \vec{\phi}' - \sum_\lambda \delta f_\lambda \vec{L}_{\lambda}\tr - \ln \vec{p}^{\mathrm{eq}} \right)\tr \vec{p}^{\mathrm{eq}'} \\ & =
	\Phi_{\mathrm{eq}'} + \left( - \sum_\lambda \delta f_\lambda \vec{L}_{\lambda}\tr - \ln \frac{\vec{p}^{\mathrm{eq}}}{\vec{p}^{\mathrm{eq}'}} \right)\tr \vec{p}^{\mathrm{eq}'} =
	\Phi_{\mathrm{eq}'} - \sum_\lambda \delta f_\lambda \vec{L}_{\lambda}\tr \vec{p}^{\mathrm{eq}'} + D_{\mathrm{KL}} \big( p^{\mathrm{eq}'} \big\| p^\mathrm{eq} \big)\,.
	\end{aligned}
\end{equation}
%Expanding the old equilibrium as $\vec{p}^{\mathrm{eq}'}= \vec{p}^\mathrm{eq} + \delta\vec{p}  + O(\abs{\delta \vec{f}}^2)$, %where the upper index refers to the order of the expansion in terms of $\delta \vec{f}$ 
%we obtain the corrections (written in components for clarity, averages are taken with respect to $\vec{p}^\mathrm{eq}$).
By solving the integral in Eq.~\eqref{eq:prob_response_db_time} with $\delta \vec{\phi}(s)=\theta(s) \sum_\lambda \delta f_{\lambda} \vec{L}_{\lambda}\tr$ one obtains the total probability deviation
\begin{align}
    \Delta p_n &\coloneq \lim_{t\to \infty} \delta p_n(t)= \sum_\lambda \delta f_\lambda \left(\avg{L_\lambda}_\mathrm{eq} - L_{\lambda n}\right)p^\mathrm{\mathrm{eq}}_{ n}\,.
\end{align}
Using this expression, and the conservation of probability, we can rewrite the total dissipation \eqref{eq:epdkl} as
\begin{align}
    \Sigma &\approx   \frac{1}{2}\avg{\left(\frac{(\Delta\vec{p})}{\vec{p}^\mathrm{eq}}\right)^2}_\mathrm{eq}
    = \frac{1}{2} \sum_{\lambda,\lambda'} \delta f_\lambda \delta f_{\lambda'} \left(\avg{L_\lambda L_{\lambda'}}-\avg{L_\lambda}\avg{L_{\lambda'}}\right) \, ,
\end{align}
up to third order corrections.
By recognizing the last term on the right hand side as the covariance matrix of the conserved extensive quantities $\vec{L}_\lambda$, we immediately see that the matrix 
\begin{align}
    \frac{\partial^2\Sigma}{\partial f_\lambda \partial f_{\lambda'}}  =-\frac{\partial^2\avg{\Phi}_\mathrm{eq}}{\partial f_\lambda \partial f_{\lambda'}}  = \frac{1}{2}\left(\avg{\vec{L}_{\lambda}\vec{L}_{\lambda'}} - \avg{\vec{L}_{\lambda}}\avg{\vec{L}_{\lambda'}}\right)
\end{align} is positive semi-definite.
The first equality holds because of \eqref{eq:epdkl}.
In the thermodynamic limit, this relations entail the so-called \emph{thermodynamic stability conditions}.
These constraints indeed follow from the fact that the the principal minors of $\partial_\lambda \partial_{\lambda'}\avg{\Phi}^\mathrm{eq}$---which corresponds to quantities such as heat capacity and compressibility---inherit the negative semi-definiteness.

In the following section, we develop the above treatement in the more general setting in which nonconservative forces and arbitrary driving protocols are present.

\subsection{EPR of periodic steady states}
\label{sec:quadraticepr_fields}

We here aim at expressing the EPR as bilinear form in terms of the protocol's Fourier amplitudes, since this provides a useful geometric characterization of the dissipation due to periodic protocols.

We start our derivation by linearizing the edge currents appearing in the EPR~\eqref{eq:epr} in terms of the edge affinities~\eqref{eq:affities_fields},
\begin{align}
     \dot{\Sigma}(t) &\approx \avg{\delta\vec{A}(t),\delta\vec{A}(t)} \,. \label{total_epr_time_linearized}
\end{align}
This expression is valid for weak perturbations with arbitrary time-dependence, and will serve as a basis for the result presented in \S\ref{sec:finite_time_EPR}.
Focusing here on the frequency domain and applying the Fourier transform~\eqref{eq:fourier} we obtain
\begin{align}
  \dot{\Sigma}(t) &\approx 
 \int \d \omega \d \omega' \e^{\i (\omega + \omega')t}\avg{\delta \hat{\vec{A}}(\omega)^*, \delta \hat{\vec{A}}(\omega')}\,.\label{eq:epr__lowest_order}
\end{align}

We seek to express the EPR \eqref{eq:epr__lowest_order} in terms of the variations of the entropic fields $\delta \vec{f}$.
We first combine Eqs.~\eqref{eq:linearizedaffinity_result} and \eqref{eq:fourier_mclennan} to express the response of the edge affinities as $\delta \hat{\vec{A}}(\omega) = 	\mathbb{R}(\omega)\delta \hat{\vec{f}}(\omega)$, with the response matrix given by
\begin{align}
	\mathbb{R}(\omega) &\coloneq  - \mathbb{X}\tr + \mathbb{D}\tr \mathbb{A}(\omega) \avg{\mathbb{D}\tr, \mathbb{X}\tr} \, .
	\label{eq:affinity_resp_matrix}
\end{align}

Consider now the specific case of a periodic driving protocol of period $T$ containing multiple commensurate frequencies $\omega_k=\frac{2\pi k}{T}$. The system will relax to a periodic steady state, as Eq.~\eqref{eq:fourier_mclennan} shows for a protocol of the form $\delta \vec{f}(t) = \sum_k \e^{\i \omega_k t} \hat{\vec{f}}(\omega_k)$.
This periodic state is reached after the longest relaxation time of the dynamics.
The edge affinities become
\begin{align}
  \delta \hat{\vec{A}}(\omega) &= \sum_{k=-\infty}^{\infty} \delta \hat{\vec{A}}(\omega_k) \delta_{\mathrm{D}}(\omega - \omega_k)
  = \sum_{k=-\infty}^{\infty} \mathbb{R}(\omega_{k}) \delta \hat{\vec{f}}(\omega_k) \delta_{\mathrm{D}}(\omega - \omega_k) \, ,
\end{align}
and we can use this expression to characterize the lowest order contribution to the EPR~\eqref{eq:epr__lowest_order} averaged over one period
\begin{align}
	\overline{\dot{\Sigma}} & \coloneq \frac{1}{T} \int_0^T \d t \sum_{j,k} \e^{\i (\omega_j+\omega_k)t} \avg{\delta \hat{\vec{A}}(\omega_{j})^*, \delta \hat{\vec{A}}(\omega_{k})} \\
	& = \sum_{k=-\infty }^{\infty} \overline{\dot{\Sigma}}(\omega_k) \, , \label{eq:epr_bilinear}
\end{align}
with
\begin{align}
   	\overline{\dot{\Sigma}}(\omega_k) &\coloneq
    \avg{ \delta\hat{ \vec{A}}(\omega_k), \delta \hat{\vec{A}}(\omega_k) } \label{eq:epr_bilinear_affinity_freq}\\
    &= \delta \hat{\vec{f}}(\omega_k)^\dagger \avg{\mathbb{R}(\omega_k),\mathbb{R}(\omega_k)} \delta \hat{\vec{f}}(\omega_k) \; .
	\label{eq:epr_bilinear2}
\end{align}

Equation~\eqref{eq:epr_bilinear} is a first result obtained from of our approach:
the entropy production per period in the periodic steady state is written in terms of a scalar product over the space of the Fourier components of the perturbation.
Different frequencies contribute to the total EPR independently from each other, as a consequence of the linearization procedure.
This is instrumental for the next result of the paper, \textit{viz.} the construction of a generalized Onsager matrix.
Notice that the expression~\eqref{eq:epr_bilinear} is such that the average EPR of each Fourier mode is non-negative, as it is the modulus of a vector, namely $\overline{\dot{\Sigma}}(\omega_k) \geq 0$.
It follows that $\overline{\dot{\Sigma}} \geq 0$.

Another implication of Eq.~\eqref{eq:epr_bilinear}  is related to time-reversal invariance. Since the matrix product~\eqref{eq:hermitian_product_matrix} is Hermitian, we have that the entropy production corresponding to a forward protocol, $\vec{f}(t)$, coincides with that of to the time reversal protocol, $\delta \vec{f}_{\mathrm{TR}}(t)=\delta \vec{f}(-t)$.
Indeed, from the equality $\delta \vec{f}_{\mathrm{TR}}(\omega_k)~=~\int \d t \e^{\i \omega_k t} \delta \vec{f}(-t) = \delta \vec{f}(\omega_k)^{*}$, we find that
\begin{align}
  \overline{\dot{\Sigma}}_{\mathrm{TR}}(\omega_k) &= \avg{\delta \hat{\vec{A}}^*(\omega_k), \delta \hat{\vec{A}}^*(\omega_k)}\ = \overline{\dot{\Sigma}}(\omega_k)\,. \label{eq:timerev_invariance}
\end{align}
We remark that, because of the presence of equilibrium fluxes in the Hermitian product \eqref{eq:hermitian_product}, the average entropy production of a mode depends on the complete knowledge of the equilibrium rates of the reference equilibrium state.
%However, as seen in Eq.~\eqref{eq:linearizedaffinity_result}, deviations of the symmetric part of the rates from their equilibrium value do not enter in the linear response near equilibrium~\cite{baiesi2013response}.

\subsection{Response of the EPR: a Generalized Onsager Matrix}
\label{sec:generalized_onsager_fields}

We now build on Eq.~\eqref{eq:epr_bilinear} and the use of the fundamental thermodynamic forces introduced in \S\ref{sec:conservationlaws} to construct a generalized Onsager matrix.

We preliminary recall the decomposition of the intensive fields introduced in Eq.~\eqref{eq:decomposition},
\begin{align}
\mathbb{T} \delta \hat{\vec{f}}(\omega_{k}) = \left(\delta \hat{\vec{\mathcal{F}}}(\omega_{k}), \delta \hat{\vec{f}}_{\mathrm{p}}(\omega_{k}) \right) \, , \label{eq:splitdeltaf}
\end{align}
which is here expressed in the frequency domain.
The first subvector accounts for the nonconservative forces $\delta \hat{\vec{\mathcal{F}}}(\omega_k)=-\delta \hat{\vec{f}}_\mathrm{f} (\omega_k) + \mathbb{l}_\mathrm{f} \, \mathbb{l}^{-1}_\mathrm{p} \delta \hat{\vec{f}}_\mathrm{p}(\omega_k)$,
while the second for the perturbation to the potential fields, which are left invariant by $\mathbb{T}$.
The latter term is responsible for changes of the reference equilibrium.
Inserting the identity matrix written as $\mathbb{1}=\mathbb{T}^2$ to the left and to the right of the quadratic form in Eq.~\eqref{eq:epr_bilinear}, we can rewrite the EPR of a mode as
\begin{align}
  \overline{\dot{\Sigma}}(\omega_k)
                          &\coloneq
                            \begin{pmatrix}
                              \delta \hat{\vec{\mathcal{F}}}(\omega_k)\\
                              \delta \hat{\vec{f}}_{\mathrm{p}}(\omega_k)
                            \end{pmatrix}^\dagger
  \begin{pmatrix}
    \mathbb{O}_{\mathrm{ff}}(\omega_k) & \mathbb{O}_{\mathrm{fp}}(\omega_k) \\
    \mathbb{O}_{\mathrm{pf}}(\omega_k) & \mathbb{O}_{\mathrm{pp}}(\omega_k) \\
  \end{pmatrix}
  \begin{pmatrix}
    \delta \hat{\vec{\mathcal{F}}}(\omega_k)\\
    \delta \hat{\vec{f}}_{\mathrm{p}}(\omega_k)
    \end{pmatrix}\,. \label{eq:epronsager}
\end{align}
The matrix
\begin{align}
	\mathbb{O}(\omega_k) =\mathbb{T}\tr\avg{\mathbb{R}(\omega_k),\mathbb{R}(\omega_k)} \mathbb{T}
	\label{eq:onsagerdef}
\end{align}
appearing in Eq.~\eqref{eq:epronsager} is the \emph{generalized Onsager matrix} of the system.
Indeed, it characterizes how the EPR responds to the variations of the fundamental forces $\hat{\vec{\mathcal{F}}}(\omega_k)$ and the fields $\hat{\vec{f}}_{\mathrm{p}}(\omega_k)$ at frequency $\omega_{k}$.
The explicit blocks of the generalized Onsager matrix~\eqref{eq:onsagerdef} are expressed in terms of the matrices $\mathbb{R}_{\mathrm{f}}$ and $\mathbb{R}_{\mathrm{p}}$ defined by 
\begin{align}
  \mathbb{R}_{\mathrm{f}} (\omega_k) &\coloneq -\mathbb{X}_{\mathrm{f}}\tr+\mathbb{D}\tr \mathbb{A}(\omega_k) \avg{\mathbb{D}\tr,\mathbb{X}_{\mathrm{f}}\tr} \,, \label{eq:matrix_r_f_def} \\
  \mathbb{R}_{\mathrm{p}} (\omega_k) &\coloneq -\mathbb{X}_{\mathrm{p}}\tr+\mathbb{D}\tr \mathbb{A}(\omega_k) \avg{\mathbb{D}\tr,\mathbb{X}_{\mathrm{p}}\tr}\,,
  \label{eq:matrix_r_p_def} 
\end{align}
where the indices p and f selects the rows of $\mathbb{X}$ consistently with the identification of nonconservative forces and potential fields in Eq.~\eqref{eq:splitting}. 
The generalized Onsager matrix in block form reads
\begin{widetext}
\begin{align}
    \mathbb{O}(\omega_k)
&=
  \begin{pmatrix}
    -\mathbb{1} & \mathbb{0}\\
    (\mathbb{l}^{-1}_{\mathrm{p}})\tr \mathbb{l}_{\mathrm{f}}\tr &  \mathbb{1}\\
  \end{pmatrix}
  \begin{pmatrix}
    \avg{\mathbb{R}_{\mathrm{f}}, \mathbb{R}_{\mathrm{f}}} & \avg{\mathbb{R}_{\mathrm{f}} , \mathbb{R}_{\mathrm{p}}}\\
    \avg{\mathbb{R}_{\mathrm{p}},\mathbb{R}_{\mathrm{f}}} & \avg{\mathbb{R}_{\mathrm{p}} , \mathbb{R}_{\mathrm{p}}}\\
  \end{pmatrix}
  \begin{pmatrix}
    -\mathbb{1} & \mathbb{l}_{\mathrm{f}}  \mathbb{l}^{-1}_{\mathrm{p}}\\
    \mathbb{0} &  \mathbb{1}\\
  \end{pmatrix} \label{eq:onsager_blocks}\\
&=
  \begin{pmatrix}
    \avg{\mathbb{R}_{\mathrm{f}},\mathbb{R}_{\mathrm{f}}} & -\left(\avg{\mathbb{R}_{\mathrm{f}}, \mathbb{R}_{\mathrm{f}}} \mathbb{l}_{\mathrm{f}}  \mathbb{l}^{-1}_{\mathrm{p}} + \avg{\mathbb{R}_{\mathrm{f}},\mathbb{R}_{\mathrm{p}}}\right)\\
   -\left(\avg{\mathbb{R}_{\mathrm{f}}, \mathbb{R}_{\mathrm{f}}} \mathbb{l}_{\mathrm{f}}  \mathbb{l}^{-1}_{\mathrm{p}} + \avg{\mathbb{R}_{\mathrm{f}},\mathbb{R}_{\mathrm{p}}}\right)\tr & (\mathbb{l}^{-1}_{\mathrm{p}})\tr \mathbb{l}_{\mathrm{f}}\tr \avg{\mathbb{R}_{\mathrm{f}}, \mathbb{R}_{\mathrm{f}}}  \mathbb{l}_{\mathrm{f}}  \mathbb{l}^{-1}_{\mathrm{p}}
    + (\mathbb{l}^{-1}_{\mathrm{p}})\tr \mathbb{l}_{\mathrm{f}}\tr \avg{\mathbb{R}_{\mathrm{f}},\mathbb{R}_{\mathrm{p}}}
    +\avg{\mathbb{R}_{\mathrm{p}},\mathbb{R}_{\mathrm{f}}} \mathbb{l}_{\mathrm{f}}  \mathbb{l}^{-1}_{\mathrm{p}}
    +\avg{\mathbb{R}_{\mathrm{p}},\mathbb{R}_{\mathrm{p}}}
  \end{pmatrix}\,. \nonumber
\end{align}
\end{widetext}

Equation~\eqref{eq:epronsager} is a central result of this work.
It is the bilinear form that gives the EPR for each frequency mode of the protocol.
Crucially, Eq.~\eqref{eq:epronsager} discriminates the perturbations of different nature. 
The blocks corresponding to $\hat{\vec{\mathcal{F}}}(\omega_k)$ describe the response to perturbation affecting the nonconservative forces. In contrast, the blocks corresponding $\hat{\vec{f}}_{\mathrm{p}}(\omega_k)$ describe how the EPR responds to variations of the reference equilibrium. 
The generalized Onsager matrix~\eqref{eq:onsagerdef} is symmetric for each frequency as it results from applying the matrix $\mathbb{T}$ and its transpose to the symmetric matrices $\avg{\mathbb{R}(\omega_k),\mathbb{R}(\omega_k)}$.

The first diagonal block at zero frequency $\mathbb{O}_\mathrm{ff}(0)$, which we call \emph{static}, plays a special role as it gives the total dissipation for steady-state processes.
We will prove in a later subsection that the other three blocks in Eq.~\eqref{eq:epronsager} do not contribute to the dissipation in steady state, \textit{viz.} $\mathbb{O}_\mathrm{fp}(0)=\mathbb{O}_\mathrm{pf}\tr(0)$ and $\mathbb{O}_\mathrm{pp}(0)$ all vanish. 
Physically, finite values of $\delta \hat{\vec{f}}_\mathrm{p}(0)$ must be understood as adiabatic, infinitesimal changes of  the reference equilibrium, thus not affecting  the EPR in a periodic steady state.

\iffalse
A related but different generalized Onsager matrix was obtained in Ref. \cite{proesmans2019general}. The main difference is that our construction guaranties that the matrix is symmetric and that the time-reversal operation ($\omega \to -\omega$ in frequency space) leaves it invariant. At the same time, we now show that our generalized Onsager $\mathbb{O}$ does not coincide with the response function of the currents, as in the previous works \cite{brandner2016periodic, proesmans2019general}.
\fi

% io scriverei senza mezzi termini:

We remark that, in contrast to previous generalizations of the Onsager matrix, our construction guaranties that \textit{(i)} such a matrix is  symmetric, \textit{(ii)} it solely depends on the system and not on the protocol, and \textit{(iii)} the time-reversal operation ($\omega \to -\omega$ in frequency space) leaves the matrix invariant, \textit{cf.} Refs.~\cite{proesmans2019general, brandner2016periodic}.

\subsection{Response of the Currents and Conserved Quantities}
\label{sec:currents_to_intensive}

We focus now on the physical currents in the periodic steady state, $\vec{I}  =  \sum_{k} \hat{\vec{I}}(\omega_k) \e^{\i \omega_k t}$, whose variations read
\begin{align}
	\delta \vec{I}  &= \mathbb{X} \delta \vec{J} =\avg{\mathbb{X}\tr, \delta \vec{A}} \, . \label{eq:current_differential_intensive}
\end{align}
Using the response function introduced in Eq.~\eqref{eq:affinity_resp_matrix}, we readily obtain the Fourier components
\begin{align}
	\delta \hat{\vec{I}}(\omega_k)
	& = \avg{\mathbb{X}\tr, \mathbb{R}(\omega_k)} \delta \hat{\vec{f}}(\omega_k) \, .
	\label{eq:physcurrent_fields}
\end{align}
The validity of Onsager reciprocal relations is clear:
From Eq.~\eqref{eq:affinity_resp_matrix}, one immediately sees that the Onsager response matrix of the physical currents, \(\avg{\mathbb{X}\tr, \mathbb{R}(\omega_k)}\) in Eq.~\eqref{eq:physcurrent_fields}, is symmetric. 

To recover the Onsager reciprocal relation for perturbations described in terms of fundamental forces and potential fields, an appropriate linear transformation of the currents is required.
Using the decomposition in terms of the fundamental forces and potential fields, Eqs.~\eqref{eq:decomposition} and \eqref{eq:splitdeltaf}, together with Eq.~\eqref{eq:physcurrent_fields}, we obtain
\begin{align}
    \begin{pmatrix}
		\delta \hat{\vec{I}}_\mathrm{f}(\omega_k) \\
		- (\mathbb{l}_\mathrm{p}^{-1})\tr\mathbb{l}_\mathrm{f}\tr\delta\hat{\vec{I}}_\mathrm{f} - \delta\hat{\vec{I}}_\mathrm{p}(\omega_k)
	\end{pmatrix}
		& = - \mathbb{T}\tr \delta \hat{\vec{I}}(\omega_k)
		= \mathbb{T}\tr \avg{- \mathbb{X}\tr, \mathbb{R}(\omega_k)} \mathbb{T}
			\begin{pmatrix}
				\delta \hat{\vec{\mathcal{F}}}(\omega_k) \\
				\delta \hat{\vec{f}}_{\mathrm{p}}(\omega_k)
			\end{pmatrix} \, . \label{eq:physicalcurr_response}
\end{align}
Again, it is readily seen that \(\mathbb{T}\tr \avg{- \mathbb{X}\tr, \mathbb{R}(\omega_k)} \mathbb{T}\) is symmetric.

We can now characterize the response of conserved quantities.
By rewriting the balance Eq.~\eqref{eq:balance_eq_currents} in terms of the perturbations in the frequency domain, and using Eq.~\eqref{eq:physicalcurr_response}, we obtain
\begin{align}
     \i \omega_k \delta \widehat{\avg{{L}_{\lambda}(\omega_k)}}=  \vec{\ell}\tr_\lambda \avg{\mathbb{X}\tr,\mathbb{R}(\omega_k)}\mathbb{T}
     \begin{pmatrix}
      \delta \hat{\vec{\mathcal{F}}}(\omega_k) \\
      \delta \hat{\vec{f}}_\mathrm{p}(\omega_k)
     \end{pmatrix}\,.
     \label{eq:varL}
\end{align}
This relation generalizes the fluctuation--response relation \eqref{eq:gk_dynamical} to nonconservative dynamics (\(\delta \vec{\mathcal{F}} \neq 0\)).
Equation~\eqref{eq:varL} will be further generalized in the next section, where we will also account for the possibility that the observables itself depends on time, \emph{cf.} Eq.~\eqref{eq:balance_general}.

\subsection{Nonequilibrium Steady States and Relation between Response Matrices}
\label{sec:relationResponseMatrices}

We now show that the linear regime thermodynamics of nonequilibrium steady states is fully characterized by a single system-dependent matrix (the ordinary Onsager matrix), in terms of which we obtain the response of both the EPR and the physical currents.
In this regime, obtained when $\omega_k=0$, for all \(k\), the right hand side of Eq.~\eqref{eq:varL} must vanish for any perturbation $\delta \hat{\vec{f}}$.
This implies that
\begin{align}
    \vec{\ell}\tr_\lambda \avg{\mathbb{X}\tr,\mathbb{R}(0)} = \vec{0} \, .
    \label{eq:lproperty}
\end{align}
A fact already anticipated in \S\ref{sec:generalized_onsager_fields}, we now prove that $\mathbb{O}_\mathrm{pf}(0)$, $\mathbb{O}_\mathrm{fp}(0)$, and $\mathbb{O}_{\mathrm{pp}}(0)$ all vanish, as well as the corresponding blocks of the response matrix of the currents, Eq.~\eqref{eq:physicalcurr_response}.
To this end, using Eqs.~\eqref{eq:affinity_resp_matrix} and \eqref{eq:auxiliarymatrA}, we obtain the identity
\begin{align}
    \avg{\mathbb{R}(0), \mathbb{R}(0)}= \avg{-\mathbb{X}\tr, \mathbb{R}(0)} \, ,
	\label{eq:ss_identity}
\end{align}
and then we rewrite Eq.~\eqref{eq:lproperty} in matrix form, and separate the blocks corresponding to forces and potential fields,
\begin{align}
	\mathbb{0}&=\mathbb{l}_\mathrm{f}\tr\avg{\mathbb{R}_\mathrm{f}(0), \mathbb{R}(0)} + \mathbb{l}_\mathrm{p}\tr\avg{\mathbb{R}_\mathrm{p}(0), \mathbb{R}(0)} \, .
\end{align}
By multiplying both sides by the inverse of $\mathbb{l}_\mathrm{p}\tr$, we obtain
\begin{align}
     (\mathbb{l}_{\mathrm{p}}^{-1})\tr\mathbb{l}_\mathrm{f}\tr\avg{\mathbb{R}_\mathrm{f}(0), \mathbb{R}(0)} =- \avg{\mathbb{R}_\mathrm{p}(0), \mathbb{R}(0)} \label{eq:zeroing_step_2}\,.
\end{align}
Finally, using some algebra and this last equality in Eq.~\eqref{eq:onsager_blocks} (with $\omega=0$) allows us to see that all blocks but $\mathbb{O}_\mathrm{ff}(0)$ vanish.

We now turn to the response matrix of the currents appearing in Eq.~\eqref{eq:physicalcurr_response}. First, we apply the identity \eqref{eq:ss_identity} to the generalized Onsager matrix~\eqref{eq:onsagerdef}, yielding 
\begin{align}
     \mathbb{T}\tr\avg{-\mathbb{X}\tr, \mathbb{R}(0)}\mathbb{T}=\mathbb{O}(0) \,, \label{eq:ss_identity_onsager}
\end{align}
\textit{i.e.} that the generalized Onsager matrix coincides---at steady state---with the response matrix of the currents, Eq.~\eqref{eq:physicalcurr_response}.
Therefore, this latter matrix  has  only one nonvanishing block  coinciding with $\mathbb{O}_\mathrm{ff}(0)$. The equivalance of these matrices is valid solely at steady state,
\begin{align}
	\delta\hat{\vec{I}}_\mathrm{f}(0) & = \avg{-\mathbb{X}_\mathrm{f}\tr, \mathbb{R}_\mathrm{f}(0)}   \delta \hat{\vec{\mathcal{F}}}(0) =\mathbb{O}_\mathrm{ff}(0)\delta \hat{\vec{\mathcal{F}}}(0) \, .
	\label{eq:steady_linearized_current}
\end{align}
in agreement with classical treatments of the linear regime \cite{onsager1931reciprocalI}. A physical consequence of the above reasoning is that no steady current can be sustained by just shifting the equilibrium state by $\delta \hat{\vec{f}}_\mathrm{p}(0)$. 
This also clarifies that the response matrix of the currents, Eq.~\eqref{eq:physicalcurr_response}, does not coincide with the generalized Onsager matrix~\eqref{eq:onsagerdef} at arbitrary frequency.

We conclude this section with a simple yet important physical implication of the linear response regime of periodically driven systems:  
No current can flow against the average thermodynamic force.
Indeed, the direction of the currents is determined by the average thermodynamic force applied since Eq.~\eqref{eq:steady_linearized_current} can  be rewritten as
\begin{align}
	\overline{\vec{I}}_\mathrm{f} & \approx \delta \hat{\vec{I}}_{\mathrm{f}}(0)
	= \mathbb{O}_{\mathrm{ff}}(0) \, \delta \vec{\hat{\mathcal{F}}}(0)
	= \mathbb{O}_{\mathrm{ff}}(0) \, \overline{\delta \vec{\mathcal{F}}} \, .
	\label{eq:no_pumping_intensive}
\end{align}
In other words, for perturbations to which this theory applies, it is not possible to establish a current against a nonconservative force $\delta \vec{\mathcal{F}}$ by acting on the potential fields $\vec{f}_\mathrm{p}$. The components $\overline{\vec{I}}_\mathrm{p}$ can be obtained as linear combinations of the fundamental currents $\overline{\vec{I}}_\mathrm{f}$, as can be seen from Eq.~\eqref{eq:physicalcurr_response}.

\subsection{Linear response theory for tightly-coupled Systems}

An important case in applications is the one of \emph{tightly-coupled} systems \cite{van2005thermodynamic},  \emph{e.g.} when transport of energy can only be achieved by moving particles from one reservoir to another, as in the Example \ref{sec:ex_2}.
Mathematically, this is expressed by the existence of a conservation law $\vec{\ell}_\mathrm{t.c.}$ that satisfies
\begin{align}
    \vec{\ell}_\mathrm{t.c.}\tr\mathbb{X}=\vec{0}\,, \label{eq:tight_coupling_def}
\end{align}
that is a stronger condition compared to eq. \eqref{eq:cl_definition}. 
It implies a constraint for the instantaneous currents in the reservoirs, as it is shown by multipying the definition on the right by the probability flux $\vec{j}$
\begin{align}
   \vec{\ell}_\mathrm{t.c.}\tr\vec{I}=\vec{0} \,,\label{eq:tight_coupling_currents}
\end{align}
In the simple case of a system with two entropic fields and exchanged quantities, the above condition is equivalent to require the proportionality of the two currents. In general, eq.~\eqref{eq:tight_coupling_currents} means that the emergence of system-specific conservation laws forces different energy transduction mechanisms to happen simultaneously.
The immediate consequence of \eqref{eq:tight_coupling_def}  is that both the response matrix of the currents \eqref{eq:physcurrent_fields} and the generalized Onsager matrix \eqref{eq:onsagerdef} have $ \vec{\ell}_\mathrm{t.c.}$ as left null vector, and will thus be degenerate.

\section{General protocol involving reservoirs and system quantities}
\label{sec:generalprotocol}

We now consider the most general class of protocols,  thus acting not only on the intensive fields of the reservoirs, $\vec{f}$, but also on the extensive quantities of the system, $\mathbb{X}$.
This is motivated, for instance, by experiments in which an electric or magnetic field is used to change the energy levels of the system.
We first derive the dynamical as well as topological effects that these protocols have on the system.
We then investigate thermodynamic forces and currents,  highlighting the differences with the simplified picture presented in Sec.~\ref{sec:driving_reservoirs}.
Finally, we establish a general expression of the near equilibrium EPR that is valid at finite times, and examine its consequences for the minimum entropy production principle and for the adiabatic regime of driving.

%-------------------
\subsection{Response of Topology and Dynamics}

The additional perturbation of the system quantities is denoted by $\mathbb{X}(t) = \mathbb{X}^\mathrm{eq} + \delta \mathbb{X}(t)$, and affects the state-wise value of the conserved quantities $\mathbb{L} \mapsto \mathbb{L} + \delta \mathbb{L}$.
In fact,  adapting the definition of conservation laws and the balance equation Eqs.~\eqref{eq:conservationlaws} and \eqref{eq:cl_definition}, we get the equation to compute the change in the conservation law matrix
\begin{align}
	\mathbb{C}\tr(\mathbb{X} + \delta\mathbb{X})\tr(\mathbb{l} + \delta\mathbb{l}) = 0 \, ,
	\label{eq:pert_cl}
\end{align}
that in turn gives the balance obeyed by the perturbed conserved quantity 
\begin{align}
	\delta \mathbb{X}\tr \vec{\ell}_\lambda + \mathbb{X}\tr\delta \vec{\ell}_\lambda = \mathbb{D}\tr \delta \vec{L}_\lambda \,.
	\label{eq:pert_balance}
\end{align}
We mention that, as pointed out in Ref.~\cite{rao2018conservation},
only the \emph{non-trivial} (system-specific) conservation law vectors can depend on the protocol, giving $\delta\mathbb{l} \neq 0$, since the \emph{trivial} ones (energy conservation, \textit{etc}.) will always be independent on the instantaneous value of the systems' parameters.

Aware of the subtleties of this general type of perturbation, we now proceed to describe how the linearized dynamics is affected.
The expression of the affinities in Eq.~\eqref{eq:linearizedaffinity_result} and the corresponding master equation in Eq.~\eqref{eq:linearizedme_affinity}, must now be replaced by
\begin{equation}
	\vec{A} \approx \delta \vec{A} = - \delta \mathbb{X}\tr \vec{f} - \mathbb{X}\tr \delta \vec{f} - \mathbb{D}\tr \delta \ln \vec{p} \, ,
\end{equation}
and
\begin{equation}
	\d_t \delta \vec{p} =- \avg{\mathbb{D}\tr, \delta \mathbb{X}\tr} \vec{f} - \avg{\mathbb{D}\tr, \mathbb{X}\tr} \delta \vec{f} - \avg{\mathbb{D}\tr,\mathbb{D}\tr} \delta \ln \vec{p} \, ,
	\label{eq:ME:generalProtocol}
\end{equation}
respectively.
In contrast, Eqs.~\eqref{eq:generalaffinities} and \eqref{eq:linearizedme_decomposed}---which feature the decomposition in forces and potential fields---remain formally correct since $\vec{\mathcal{F}}^\mathrm{eq} = \vec{0}$.
However, the perturbations of the Massieu potential now read
\begin{align}
	\delta \vec{\mathcal{F}} & = - \delta \vec{f}_{\mathrm{f}} + \mathbb{l}_{\mathrm{f}} \, \mathbb{l}_{\mathrm{p}}^{-1} \delta \vec{f}_{\mathrm{p}} + \delta(\mathbb{l}_{\mathrm{f}} \, \mathbb{l}_{\mathrm{p}}^{-1}) \vec{f}_{\mathrm{p}} \label{eq:most_gen_force} \\
	\delta \vec{\phi} & =
	- \mathbb{L} \, (\mathbb{l}^{-1}_{\mathrm{p}}) \, \delta \vec{f}^{\vphantom{-1}}_{\mathrm{p}}
	- \delta\mathbb{L} \, (\mathbb{l}^{-1}_{\mathrm{p}}) \, \vec{f}^{\vphantom{-1}}_{\mathrm{p}}
	- \mathbb{L} \, \delta(\mathbb{l}^{-1}_{\mathrm{p}}) \, \vec{f}^{\vphantom{-1}}_{\mathrm{p}} \, ,
\end{align}
\textit{cf.} Eq.~\eqref{eq:deltaphi:fieldsPert}.
Note that the nonconservative forces are affected by perturbations of system quantities solely through the conservation law vector (the last term in Eq.~\eqref{eq:most_gen_force}), see Example \ref{sec:ex_2}.

\subsection{Current response}
\label{sec:currents_general}
The response of the physical currents changes when the perturbation acts both on $\mathbb{X}$ and $\vec{f}$ compared to the situation in Section~\ref{sec:currents_to_intensive}. 
It features time-dependent terms containing $\delta \mathbb{X} $ and $\delta \vec{f}$.
Since at the reference equilibrium $\vec{J}^{\mathrm{eq}}=0$ by definition, the first order contribution to the physical currents is
\begin{align}
  \delta \vec{I} &= \delta (\mathbb{X} \vec{J}) =  \mathbb{X} \delta \vec{J}= - \avg{\mathbb{X}\tr,\delta (\mathbb{X}\tr \vec{f}) + \mathbb{D}\tr \delta \vec{p}} \label{eq:currents_general_fields_first}\\
                 &= - \avg{\mathbb{X}\tr,\mathbb{X}\tr} \delta\vec{f} - \avg{\mathbb{X}\tr,  \delta\mathbb{X}\tr }  \vec{f}^{\mathrm{eq}} - \avg{\mathbb{X}\tr,\mathbb{D}\tr} \delta \vec{p}\,.\label{eq:currents_general_fields}
\end{align}

Comparing Eq.~\eqref{eq:currents_general_fields} to Eq.~\eqref{eq:current_differential_intensive} combined with \eqref{eq:linearizedaffinity_result} , we note that an additional term containing the change $\delta \mathbb{X}$ of exchanged quantities appears.
In addition, $\delta \mathbb{X}$ also implicitly determines the solution $\delta \vec{p}$ to the linearized master equation, Eq.~\eqref{eq:ME:generalProtocol}. 
As already seen, the situation is conceptually simpler in the language of fundamental forces and Massieu potentials of the state. 
Using the decomposition~\eqref{eq:splitting_flat}, we rewrite Eq.~\eqref{eq:currents_general_fields} as
\begin{align}
  \delta \vec{I} &=  \avg{\mathbb{X}\tr,\mathbb{X}_\mathrm{f}\tr} \delta \vec{\mathcal{F}} + \avg{\mathbb{X}\tr,\mathbb{D}\tr} \delta (\vec{\phi} - \ln \vec{p}) \label{eq:currents_general} \, ,
\end{align}
where we used $\vec{\mathcal{F}}^\mathrm{eq} = \vec{0}$.
This expression still contains $\delta \vec{p}$, which can here regarded as an implicit function of $\delta \vec{\mathcal{F}}$ and $\delta \vec{\phi}$ through the master equation~\eqref{eq:linearizedme_decomposed}.
By explicating the solution for each Fourier mode,
\begin{align}
\delta \ln \hat{\vec{p}}(\omega_k) = \mathbb{A}(\omega_k) \left( \avg{\mathbb{D}\tr, \mathbb{X}_\mathrm{f}\tr} \delta \hat{\vec{\mathcal{F}}}(\omega_k)  + \avg{\mathbb{D}\tr, \mathbb{D}\tr} \delta \hat{\vec{\phi}} (\omega_k) \right) \, , \label{FTrespGen}
\end{align}
we can rewrite the current as
\begin{align}
    \begin{pmatrix}
     \hat{\vec{I}}_\mathrm{f}(\omega_k)\\
     \hat{\vec{I}}_\mathrm{p}(\omega_k)
    \end{pmatrix}
    =
    \begin{pmatrix}
    \mathbb{Q}_{\mathrm{ff}}(\omega_k)  & \mathbb{Q}_{\mathrm{fp}}(\omega_k)\\
    \mathbb{Q}_{\mathrm{pf}}(\omega_k)  & \mathbb{Q}_{\mathrm{pp}}(\omega_k)
    \end{pmatrix}
    \begin{pmatrix}
     \delta \hat{\vec{\mathcal{F}}}(\omega_k)\\
     \delta \hat{\vec{\phi}}(\omega_k)
    \end{pmatrix} \, .
    \label{eq:curr_full_response}
\end{align}
where
\begin{align}
    \begin{pmatrix}
     \mathbb{Q}_{\mathrm{ff}} (\omega)\\
     \mathbb{Q}_{\mathrm{pf}} (\omega)
    \end{pmatrix} 
    \coloneq  
     \avg{\mathbb{X}\tr,\mathbb{R}_\mathrm{f}(\omega)}
\end{align}
 and
\begin{align}
    \begin{pmatrix}
     \mathbb{Q}_{\mathrm{fp}}(\omega) \\
     \mathbb{Q}_{\mathrm{pp}}(\omega)
    \end{pmatrix} 
    &\coloneq  
     \avg{\mathbb{X}\tr,\mathbb{D}\tr}\left\{\mathbb{1}- \mathbb{A}(\omega) \avg{\mathbb{D}\tr, \mathbb{D}\tr} \right\} \label{eq:Qfp1}\\
    &= \i \omega \avg{\mathbb{X}\tr,\mathbb{D}\tr} \mathbb{A}(\omega) \mathbb{P}\label{eq:Qfp}
\end{align}
We obtained the equality \eqref{eq:Qfp} by writing  $\mathbb{1}=\mathbb{A}(\omega)\mathbb{A}^{-1}(\omega)$ and using the definition \eqref{eq:auxiliarymatrA}.

Since the number of potential fields (the dimension of $\vec{I}_\mathrm{p}$) is different from the number of nodes over which the Massieu potential $\vec{\phi}$ is defined, the response matrix \(\mathbb{Q}\) is not square in general, and Onsager symmetry for the current response coefficients is lost.
However, the symmetry is restored at $\omega=0$, when solely \(\mathbb{Q}_{\mathrm{ff}}(\omega) = \avg{\mathbb{R}_\mathrm{f}(0),\mathbb{R}_\mathrm{f}(0)} = \mathbb{Q}_{\mathrm{ff}}\tr(\omega)\) does not vanish.
Indeed, $\mathbb{Q}_{\mathrm{fp}}(0)=\mathbb{0}$ and $\mathbb{Q}_{\mathrm{pp}}(0)=\mathbb{0}$ from the definition \eqref{eq:Qfp}, while $\mathbb{Q}_\mathrm{pf}(0)=\mathbb{0}$ is a consequence of Eq. \eqref{eq:lproperty}.

We conclude this subsection with a remark concerning the possibility of \emph{pumping}, \textit{i.e.} inducing a current by changing the Massieu potentials $\vec{\phi}$ in absence of net time-averaged forces ($\delta \hat{\vec{\mathcal{F}}}(0)=\vec{0}$). 
Since $\mathbb{Q}_{\mathrm{fp}}(0)=\mathbb{0}$ and $\mathbb{Q}_{\mathrm{pp}}(0)=\mathbb{0}$, the zero-frequency response to perturbation of the potentials $\vec{\phi}$ reduces to zero.
This confirms the intuition gained in the discussion of Eq.~\eqref{eq:no_pumping_intensive}:
no pumping is possible in the realm of linear stochastic thermodynamics.
This extends the original \emph{no-pumping theorem} \cite{chernyak2008pumping, rahav2008directed} to cases in which the nonconservative forces are instantaneously different from zero, complenentig previous results on nonequilibrium systems with a specific form of the transition rates~\cite{maes2010general}. The is consistent with previous observations on driving-induced net average  currents in systems governed by time-dependent master equations \cite{sinitsyn2007universal, forastiere2020strong}, where the observed effect is nonlinear in the driving amplitude.

\subsection{Response of Conserved Quantities}
\label{sec:balanceII}

The analysis of the response of the conserved quantities in \S~\ref{sec:currents_to_intensive} is here repeated for the case of general perturbations.
This time, one needs to introduce the values of the conserved quantities associated to the reference equilibrium state, that we denote by  $\vec{L}_\lambda^\mathrm{eq}$.
The overall change of the average conserved quantities can be then properly defined,
\begin{align}
  \delta \avg{L_\lambda(t)}
  &\coloneq \vec{L}_{\lambda}\tr(t) \vec{p}(t) - (\vec{L}_{\lambda}^\mathrm{eq})\tr \vec{p}^\mathrm{eq}  \, ,
\end{align}
and, upon linearizing and Fourier transforming its time derivative, with the help of Eq.~\eqref{eq:me}, we find 
\begin{align}
   \i \omega  \delta \widehat{\avg{L_\lambda(\omega)}}& = 
  \i \omega {\avg{\delta \hat{L}_\lambda(\omega)}}_{\mathrm{eq}} + (\vec{L}_{\lambda}^{\mathrm{eq}}) \tr \mathbb{D}\delta \hat{\vec{J}} (\omega)
 \label{eq:gk_most_general}
\end{align}
where $\avg{\delta \hat{L}_\lambda(\omega)}_{\mathrm{eq}}\coloneq \delta \hat{\vec{L}}_{\lambda}\tr(\omega) \vec{p}^{\mathrm{eq}} $.  In the periodic steady state, using eqs.~\eqref{eq:conservationlaws}, \eqref{eq:physcurrentdef}, and \eqref{eq:curr_full_response}  we find
\begin{align}
 \i \omega_k\,  \delta \widehat{\avg{L_\lambda(\omega_k)}}& = 
  \i \omega_k\, {\avg{\delta \hat{L}_\lambda(\omega_k)}}_{\mathrm{eq}}
  + \vec{\ell}\tr \mathbb{Q}(\omega_k) 
  \begin{pmatrix}
   \delta \hat{\vec{\mathcal{F}}}(\omega_k)\\
   \delta \hat{\vec{\phi}}(\omega_k)
  \end{pmatrix}\,.
    \label{eq:balance_general}
\end{align}
This important relation is the general version of the Agarwal--Green--Kubo response formula when the conservation laws are chosen as observables and the perturbation can act both on the reservoirs and on the system quantities. 

The generalized version of the Agarwal--Green--Kubo relations~\eqref{eq:gk_dynamical} are recovered in detailed balanced systems, for which $\delta \vec{\mathcal{F}}=\vec{0}$.
Indeed, after expliciting $\mathbb{Q}$, Eq.~\eqref{eq:Qfp}, with the help of Eq.~\eqref{eq:conservationlaws}, Eq.~\eqref{eq:balance_general} becomes 
\begin{align}
  &\delta \widehat{\avg{{L}_\lambda(\omega_k)}} =  \avg{\delta \hat{L}_{\lambda}(\omega)}_{\mathrm{eq}}
+ \vec{L}_\lambda\tr \avg{\mathbb{D}\tr,\mathbb{D}\tr}\mathbb{A}(\omega_k) \vec{p}^{\mathrm{eq}} \delta \hat{\vec{\phi}}(\omega_k)  \,. \label{eq:gk_fourier}
\end{align}

By anti-transforming and using Eq.~\eqref{eq:observable_evoluted}, we recover the expression of the response coefficients in terms of integrals of equilibrium correlators,
\begin{align}
  \delta \avg{L_\lambda(t)}= \avg{\delta \vec{L}_\lambda(t)}_{\mathrm{eq}} + \int_{-\infty}^{+\infty} \d \tau \avg{ \vec{L}_\lambda(t-\tau)\tr \frac{\avg{\mathbb{D}\tr,\mathbb{D}\tr}}{\mathbb{P}} \delta \vec{\phi}(\tau)}_{\mathrm{eq}} \, .
    \label{eq:gk}
\end{align}
The first contribution  accounts for the explicit time-dependency of $\delta \vec{L}_{\lambda}$ on the driving protocol, and was assumed to be zero in the derivation of Eq.~\eqref{eq:gk_dynamical}. 

\subsection{Response of the EPR}
\label{sec:stability}

We complete our analysis by expressing the perturbations of the EPR.
Remarkably, two distinct decomposition of the EPR are possible, and they shed light on different aspects of the dynamics near equilibrium.
These decompositions constitute the main result of this paper.
The first decomposition is presented in \S\ref{sec:general_epr_dec_periodic} and holds for periodic protocols in the periodic steady state.
It provides a generalization of Eqs.~\eqref{eq:epronsager} and \eqref{eq:onsager_blocks}. %in which the fundamental control parameters that rule the dissipation are identified by exploiting the conservation laws.
The second decomposition, introduced in \S\ref{sec:finite_time_EPR}, holds at finite time for an arbitrary protocol.
It tells apart the contribution to the dissipation due to nonconservative forces from that caused by the time-dependent protocol and transient relaxations.
This latter decomposition is related to the minimum entropy production theorem of Prigogine~\cite{glansdorff1971thermodynamic}, which we recover as a corollary.

\subsubsection{Time-averaged EPR decomposition}
\label{sec:general_epr_dec_periodic}

For systems in the periodic steady state caused by generic protocols, the EPR~\eqref{eq:epr_bilinear_affinity_freq} retains its validity.
By expanding the edge affinity vector in terms of the variations of the nonconservative forces $\delta \hat{\vec{\mathcal{F}}}$ and of the Massieu potentials $\delta \hat{\vec{\phi}}$, Eq.~\eqref{eq:generalaffinities}, we obtain the average EPR per period of a single mode
\begin{widetext}
\begin{align}
\overline{\dot{\Sigma}}(\omega_k) &= 
\begin{pmatrix}
 \delta \hat{\vec{\mathcal{F}}} (\omega_k)\\
 \delta \hat{\vec{\phi}}(\omega_k)
\end{pmatrix}\tr
    \tilde{\mathbb{O}}(\omega_k)
\begin{pmatrix}
 \delta \hat{\vec{\mathcal{F}}} (\omega_k)\\
 \delta \hat{\vec{\phi}}(\omega_k)
\end{pmatrix}\label{eq:epr_most_general}
\end{align}
with the Onsager matrix defined as
\begin{align}
\tilde{\mathbb{O}}(\omega_k) &\coloneq 
\begin{pmatrix}
      \mathbb{O}_\mathrm{ff}(\omega_k)
     &  \omega_k^2 \avg{\mathbb{X}\tr_\mathrm{f}, \mathbb{D}}\mathbb{P} \mathbb{A}(-\omega_k) \mathbb{A}(\omega_k)\mathbb{P}\\
   \omega_k^2 \mathbb{P}\mathbb{A}(-\omega_k)\mathbb{A}(\omega_k)\mathbb{P} \avg{\mathbb{D}\tr,\mathbb{X}\tr_\mathrm{f}} & \omega_k^2 \mathbb{P}\mathbb{A}(-\omega_k)\avg{\mathbb{D}\tr,\mathbb{D}\tr}\mathbb{A}(\omega_k)\mathbb{P} 
    \end{pmatrix} \, ,
    \label{eq:general_final_onsager}
\end{align}
\end{widetext}
To characterize the latter $(\abs{\mathcal{Y}} -\abs{\Lambda}+\abs{\mathcal{V}})\times(\abs{\mathcal{Y}} -\abs{\Lambda}+\abs{\mathcal{V}})$-matrix, we also used the response term \(\delta \ln \hat{\vec{p}}(\omega_k)\) in Eq.~\eqref{FTrespGen}, as well as the property of the matrix $\mathbb{A}(\omega)$ already used in \eqref{eq:Qfp}, \textit{viz.} $\mathbb{1}-\mathbb{A}(\omega)\avg{\mathbb{D}\tr,\mathbb{D}\tr}=\i\omega \mathbb{A}\mathbb{P}$.
Equation~\eqref{eq:epr_most_general} gives the most general formulation of the time-averaged EPR of each frequency mode in the periodic steady state.
It is expressed as a quadratic form of the nonconservative forces and the Massieu potentials.
A different representation of the EPR could be obtained choosing the variations $\delta \mathbb{X}$ as control parameters.
However, Eq.~\eqref{eq:epr_most_general} is the representation with the minimal number of ``generalized thermodynamic forces'':
the nonconservative forces and variations of the Massieu potentials.
In fact, if we subtract the dimension of the vectors featuring \eqref{eq:epr_most_general}, namely $\abs{\mathcal{Y}}-\abs{\Lambda}+\abs{\mathcal{V}}$, from the number of independent entries of the matrix $\mathbb{X}$, $\abs{\mathcal{Y}}+\abs{\mathcal{K}}(\abs{\mathcal{V}}-1)$, we  reduce the dimensionality by $\Delta = \abs{\Lambda} - \abs{\mathcal{K}} +\abs{\mathcal{V}}(\abs{\mathcal{K}}-1)\geq0$, with the equality attained by unconditionally detailed balanced systems with a single conserved quantity. 
We emphasize that such a complexity reduction is obtained thanks to systematic identification and use of the conservation laws of the system.

We finally notice that the generalized Onsager matrix \eqref{eq:general_final_onsager} is symmetric at every frequency, and that all blocks except the top-left one, \(\mathbb{O}_\mathrm{ff}(\omega_k)\) vanish identically at $\omega=0$.
It reduces to equation \eqref{eq:onsager_blocks} when the protocol does not affect the system quantities.

\subsubsection{Finite-time EPR decomposition}
\label{sec:finite_time_EPR}

In contrast to the previous characterization of the EPR, the decomposition of the EPR that we describe here is valid at finite times and not only in the periodic steady state.
It is expressed in terms of the instantaneous steady-state distribution $\vec{p}^{\mathrm{ss}}(t)$, which is obtained from the linearized master equation, Eqs.~\eqref{eq:linearizedme} and \eqref{eq:linearizedme_decomposed}, as
\begin{align}
	\avg{\mathbb{D}\tr, \delta \vec{A}^\mathrm{ss}} =
	\avg{\mathbb{D}\tr, \mathbb{X}_{\mathrm{f}}\tr \delta\vec{\mathcal{F}} + \mathbb{D}\tr (\delta\vec{\phi} - \delta\ln \vec{p}^\mathrm{ss})} = \vec{0} \, .
	\label{eq:steadystatedeviations}
\end{align}
Using this relation and the following identity (see the expression of the affinity~\eqref{eq:generalaffinities}),
\begin{align}
    \delta \vec{A} + \mathbb{D}\tr \delta\ln \vec{p}
	= \mathbb{X}_{\mathrm{f}}\tr \delta\vec{\mathcal{F}} + \mathbb{D}\tr \delta\vec{\phi}
	= \delta \vec{A}^\mathrm{ss} + \mathbb{D}\tr \delta\ln \vec{p}^\mathrm{ss} \, ,
\end{align}
we can rewrite the EPR \eqref{total_epr_time_linearized} as
\begin{align}
    \dot{\Sigma} &\approx \avg{\delta \vec{A}^\mathrm{ss},\delta \vec{A}^\mathrm{ss}} + (\delta \ln \vec{p} - \delta\ln \vec{p}^\mathrm{ss})\tr \avg{\mathbb{D}\tr,\mathbb{D}\tr} (\delta \ln \vec{p} - \delta \ln \vec{p}^\mathrm{ss}) \, .
	\label{eq:epr_orthogonalized}
\end{align}
The steady-state affinities \(\delta\vec{A}^{\mathrm{ss}}\) appearing in the first term on the right hand side can be expressed in terms of the sole nonconservative forces \(\delta\vec{\mathcal{F}}\).
Indeed, using the solution of Eq.~\eqref{eq:steadystatedeviations} obtained through the pseudoinverse matrix $\avg{\mathbb{D}\tr,\mathbb{D}\tr}^{-1} = \lim_{\omega\to 0}\mathbb{A}(\omega)$,
\begin{align}
    \delta\ln \vec{p}^\mathrm{ss}
	= \avg{\mathbb{D}\tr,\mathbb{D}\tr}^{-1} \avg{\mathbb{D}\tr,\mathbb{X}\tr} \delta\vec{\mathcal{F}} + \delta\vec{\phi} \, ,
\end{align}
we can rewrite the steady-state affinities (see Eq.~\eqref{eq:generalaffinities}) as
\begin{align}
	\delta\vec{A}^\mathrm{ss} = (\mathbb{X}\tr - \mathbb{D}\tr \avg{\mathbb{D}\tr,\mathbb{D}\tr}^{-1}\avg{\mathbb{D}\tr,\mathbb{X}\tr}) \delta\vec{\mathcal{F}} \, .
\end{align}
We complete our derivation by substituting this expression and the definition of generalized Onsager matrix, Eq.~\eqref{eq:onsagerdef}, into the EPR~\eqref{eq:epr_orthogonalized},
\begin{align}
	\dot{\Sigma} &\approx \delta\vec{\mathcal{F}}\tr \mathbb{O}_{\mathrm{ff}}(0) \delta\vec{\mathcal{F}}
	+ (\delta \ln \vec{p} - \delta\ln \vec{p}^\mathrm{ss})\tr \avg{\mathbb{D}\tr,\mathbb{D}\tr} (\delta \ln \vec{p} - \delta \ln \vec{p}^\mathrm{ss}) \, .
	\label{eq:sstransient}
\end{align}
This decomposition of the dissipation rate into two separately positive quadratic forms is a main result of this paper.
The first term describes the dissipation due to the currents generated by the nonconservative forces, and it features the block of the Onsager matrix $\mathbb{O}_{\mathrm{ff}}(0)$.
It vanishes solely for unconditionally detailed balanced systems or for detailed-balanced driving protocols, when no \(\delta\vec{\mathcal{F}}\) are present or \(\delta\vec{\mathcal{F}}\) vanish at all times, respectively.
In contrast, the second term accounts for the dissipation caused by transient effects, \textit{i.e.} instantaneous deviations from the steady state.

\paragraph*{Remark} The decomposition of the entropy production in Eq.~\eqref{eq:sstransient} can be specialized for  a detailed-balanced protocol $\delta \vec{\mathcal{F}}=\vec{0}$. In this case $\delta \ln\vec{p}^\mathrm{ss}=\delta \vec{\phi}$, and if we substitute $\delta \ln \vec{p}$ from Eq.~\eqref{eq:linearizedme_decomposed} in Eq.~\eqref{eq:sstransient},  we obtain 
\begin{align}
    \dot{\Sigma}&\approx  \d_t \delta\vec{p}\tr\avg{\mathbb{D}\tr, \mathbb{D}\tr}^{-1} \d_t \delta\vec{p} \,. \label{eq:dbdriving_epr}
\end{align}
This can be intepreted as the entropy production of a system that relaxing toward an equilibrium state that is changing in time with a finite speed. In this sense, it extends the result on the EPR of relaxation obtained in Section \ref{sec:relaxation}, that holds only for instantaneous switching protocols, to cases in which the integral \eqref{eq:prob_response_db_time} cannot be computed explicitly. $\square$

\paragraph*{Remark} An interesting limit is the one of adiabatic driving, that is obtained when the Massieu potential has the form $\vec{\phi}=\vec{\phi}(kt)$, with the driving speed parameter $k$  much smaller than the intrinsic timescales of the system. The probability distribution is assumed to follow the driving as $\delta \vec{p}\approx\delta \vec{p}^{(0)}(kt) + O(k)$. By inserting this adiabatic ansatz in the master equation~\eqref{eq:linearizedme_decomposed} we find $\delta \ln \vec{p}=\delta \vec{\phi}(kt)  +O(k)$. 
Substituting back in~\eqref{eq:dbdriving_epr} we find an adiabatic expansion of the EPR for a detailed-balance protocol that allows us  to  define a thermodynamic length  when both systems quantities and  entropic fields can vary, analogously to the friction matrix~\cite{sivak2012thermodynamic}, using the matrix $\avg{\mathbb{D}\tr, \mathbb{D}\tr}^{-1}$. $\square$

We conclude this section by discussing an important implication of Eq.~\eqref{eq:sstransient}. 
Since both terms in Eq.~\eqref{eq:sstransient} are separately positive-definite, and since the sole dependence on the instantaneous state comes from the second term, the EPR is minimized by the steady-state distribution, \textit{i.e.} when $\delta \vec{p}(t) = \delta \vec{p}^{\mathrm{ss}}$ at all times.
This is the \emph{minimum entropy production principle} derived  by Prigogine, valid in the phenomenological framework of linear irreversible thermodynamics ~\cite{glansdorff1971thermodynamic, keizer1974qualms, hunt1987dissipation, ross2005exact}.
%Our approach has the virtue to emphasize the relation to the phenomenological approach because the fundamental forces are the analogous of the macroscopic forces. 
A related result is the fact that the EPR is a Lyapunov function of the relaxation dynamics.
This fact was derived long ago \cite{schnakenberg1976network,jiu1984stability,mou1986stochastic} for stochastic dynamics, and in our formalism is readily recovered. 
Indeed, starting from Eq.~\eqref{total_epr_time_linearized}
the derivative of the EPR reads
\begin{align}
  \d_t \dot{\Sigma} &\approx \d_t \avg{\delta \vec{A}, \delta \vec{A}} = 2 \avg{\delta \vec{A}, \d_t \delta \vec{A}} \;. \label{eq:lyapunov_halfway}
\end{align}
Taking the time derivative of \eqref{eq:linearizedaffinity_result} in absence of time-dependent driving and using the linearized master equation~\eqref{eq:linearizedme_affinity} gives us
\begin{align}
  \d_t \delta \vec{A} &= - \mathbb{D}\tr \mathbb{P}^{-1} \avg{\mathbb{D}\tr, \delta \vec{A}} \, .
\end{align}
Substituting this expression back into Eq.~\eqref{eq:lyapunov_halfway}, we find 
\begin{align}
 \d_t \dot{\Sigma} \approx - 2 \avg{\delta \vec{A}, \mathbb{D}\tr} \mathbb{P}^{-1} \avg{\mathbb{D}\tr, \delta \vec{A}} \approx -2 \mathcal{I}[\vec{p}(t)] \leq 0 \,. \label{eq:lyapunov}
\end{align}
where the last equality uses the definition of Fisher information \cite{cover1991information} $\mathcal{I}[\vec{p}(t)]  \coloneq \avg{(\d_t\ln\vec{p})^2}$ computed up to linear orders  using Eq.~\eqref{eq:linearizedme_affinity}.

\begin{table}%[htb]
  \begin{tabular}{cc}
    \toprule
    Symbol &  Object \\
    \midrule
    $\mathcal{V}=\{n\}$ & Mesoscopic states (nodes) \\
    $\mathcal{E}^{\pm}=\{\pm e\}$ & Transition mechanisms (edges) \\
    $\mathcal{K}=\{\kappa\}$ & Exchanged physical quantities  \\
    $\mathcal{P}=\{\rho\}$ & External physical reservoirs \\
    $\mathcal{Y}=\{(\rho,\kappa)\}$ & Generalized reservoirs  \\
    $\mathcal{A}=\{\alpha\}$ & Fundamental cycles \\
    $\Lambda=\{\lambda\}$ & Conservation laws \\
    $\vec{p}=(p_{n})$ & Probability vector \\
    $\vec{j}=(j_{e})$ & Fluxes \\
    $\vec{w}=(w_{e})$ & Transition rates \\
    $\vec{J}=(J_{e})$ & Net probability current  \\
    $\vec{I}=(I_{y})$ & Net physical current\\
    $\vec{S}=(S_{n})$ & Internal (mesoscopic) entropy \\
    $\vec{f}=(f_{y})$ & Intensive entropic fields \\
    $\vec{A}=(A_{e})$ & Edge affinity \\
    $\mathbb{D}=(\mathbb{D}_{ne})$ & Incidence matrix \\
    $\mathbb{C}=(\mathbb{C}_{e\alpha})$ & Topological cycles \\
    $\mathbb{W}=(\sum_{e}  \mathbb{D}_{en} w_e \delta_{n' \mathrm{o}(e) })$ & Rate matrix \\
    $\mathbb{P}=(p_n^\mathrm{eq}\delta_{nn'})$ & Equilibrium distribution matrix \\
    $\mathbb{M}=(\mathbb{M}_{y\alpha})$ & Physical topology \\
    $\mathbb{l}=(\mathbb{l}_{e\lambda})$ & Conservation laws \\
    $\mathbb{L}=(\mathbb{L}_{n \lambda})$ & Conserved quantities \\
    $\mathbb{X}=(\mathbb{X}_{ye})$ & Extensive exchanged quantity \\
    $\mathbb{R}=(\mathbb{R}_{ye})$ & Current-to-affinity response  \\
    $\mathbb{O}=(\mathbb{O}_{yy'})$ & Generalized Onsager matrix  \\
                                    & (perturbing reservoirs only)\\
    $\tilde{\mathbb{O}} $ & Generalized Onsager matrix \\
                         & (general case)\\
    $\mathbb{Q} $ & Currents response matrix \\
                         & (general case)\\
    \bottomrule
  \end{tabular}
  \caption{ Table of the main quantities used in the manuscript. We use capital italic for sets, bold italic for vectors and blackboard bold for matrices. Cardinality of a set is denoted using vertical bars, \emph{e.g.} $\left|\mathcal{E}^{+}\right|$.}\label{tab:notationII}
\end{table}

\section{Examples}
\label{sec:examples}

In this Section we illustrate our theory on some examples that help elucidating the main results.

\subsection{Quantum dot in contact with a time-dependent reservoir}
\label{sec:ex_1}

\begin{figure}
    \subfloat[Schematic representation]{
      \def\svgwidth{0.2\textwidth}
      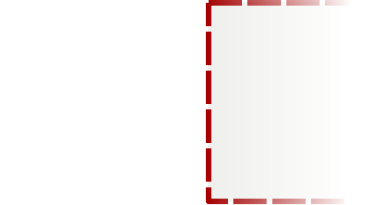}
      \quad
    \subfloat[Network]{
      \def\svgwidth{0.2\textwidth}
      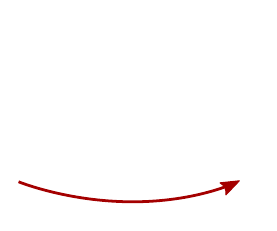}
    \caption{Single-level QD connected to a reservoir of particles.}
    \label{fig:singleqd}
\end{figure}

The simplest system we consider is a single quantum dot working in the Coulomb blockade regime, where only a single electronic state with  energy $\epsilon$ is available, Fig.~\ref{fig:singleqd}a.
The quantum dot is connected to a lead with inverse temperature $\beta$ and chemical potential $\mu$.
We call $p(t)$ the probability of occupation at time $t$, namely $p(t)=P(n=1,t)$, with $n$ being the occupation number of the quantum dot.
This probability evolves according to the master equation
\begin{align}
    \d_t p(t) = -(w_{-1}+w_{+1})p(t)+w_{+1}\,,
\end{align}
where $w_{\pm 1}$ are the transition rates of the charge and discharge events.
The local detailed balance \eqref{eq:ldb}  takes the form
\begin{align}
\frac{w_{+1}}{w_{-1}}=\e^{-\beta(t) (\epsilon - \mu)}
\end{align}
with $(\epsilon - \mu) $. The equilibrium value of the occupancy is $p^\mathrm{eq}=\e^{-\beta (\epsilon - \mu)} (1+\e^{-\beta (\epsilon - \mu)})^{-1}$.
The occupation number is $n\in\{0,1\}$, and the only forward transition is labelled $+1$.
The network representation of the system is shown in Fig.~\ref{fig:singleqd}b.
Its incidence matrix reads
\begin{align}
\mathbb{D}= \, \kbordermatrix{
  & +1  \cr
  0& -1 \cr
  1& 1 }\,.
\end{align}
This system is unconditionally detailed-balanced because of the absence fundamental nonconservative forces.
The matrix of the exchanged quantities can be written as 
\begin{align}
  \mathbb{X}\tr
  & = \, \kbordermatrix{
     & E_{\mathrm{l}}& N_{\mathrm{l}} \\
    +1 & \epsilon & 1 }\,.
\end{align}
using the fact that during the transition path we are simultaneously exchanging both energy and particles. We see immediately that there is a tight-coupling conservation law $\vec{\ell}_\mathrm{t.c.}\tr=(1, -\epsilon)$ such that $\vec{\ell}_\mathrm{t.c.}\tr\mathbb{X}=\vec{0}$, since matter and energy are exchanged by the same physical mechanism. The entropic fields that characterize the reservoirs are $\vec{f}=(\beta, - \mu \beta)$. 
The equilibrium flux reduces to a single scalar value
\begin{align}
  j^{\mathrm{eq}}&=   w_{-1}^{\mathrm{eq}} p^{\mathrm{eq}} = w_{+1}^{\mathrm{eq}}(1-p^\mathrm{eq})\,.
\end{align}
The response matrices are constructed starting from the auxiliary matrices
  \begin{align}
  \i \omega \mathbb{P} + \avg{\mathbb{D}\tr, \mathbb{D}\tr}&=
\begin{pmatrix}
  \i \omega p^{\mathrm{eq}} +  j^{\mathrm{eq}} & -  j^{\mathrm{eq}} \\
   -  j^{\mathrm{eq}} & \i \omega (1-p^{\mathrm{eq}}) +  j^{\mathrm{eq}} \\
 \end{pmatrix}\\
  \avg{\mathbb{D}\tr,\mathbb{X}\tr}&=
\begin{pmatrix}
  -\epsilon j^{\mathrm{eq}}  & -j^{\mathrm{eq}} \\
  \epsilon j^{\mathrm{eq}} & j^{\mathrm{eq}}\\
 \end{pmatrix}\,.
\end{align}

\begin{figure}
    \begin{tikzpicture}
      \begin{axis}[
        width=0.5\textwidth,
        height=0.33\textwidth,
        smooth,
        thick,
        xlabel=$\Omega$,
        domain=0:8,
        cycle list/Set2-3] %see p. 221 of pgfplots manual
        \addplot (\x, { 0.1 *  \x^2 / ( (1 + exp(-0.1)) *  ((1 + 1 * exp(0.1))^2 + \x^2 ) )}) node[below] {$\Re Y$};
        \addplot (\x, { 0.1 * exp(-0.1) *  \x / ( (1 + exp(-0.1))^2  + exp(-0.2) * \x^2) )  )
        }) node[above] {$\Im Y$};
        \addplot (\x, {0.01 * \x^2 / (  (1+exp(-0.1)) * ((1+ exp(0.1))^2 + \x^2) )  )
        }) node[above left] {$(\epsilon - \mu)^2 \mathbb{O}_{2,2}(\Omega)$} ;
      \end{axis}
    \end{tikzpicture}
 \caption{ Current response function and entropy production response function to a perturbation $\delta \beta$ of the inverse temperature.
Parameters: $w^{\mathrm{eq}}_{+1}=1, \epsilon - \mu=0.1, \beta=1$. The real part of the stochastic admittance is related to dissipation and both vanish for $\Omega~=~0$. }
\label{fig:stochastic_admittance}
\end{figure}
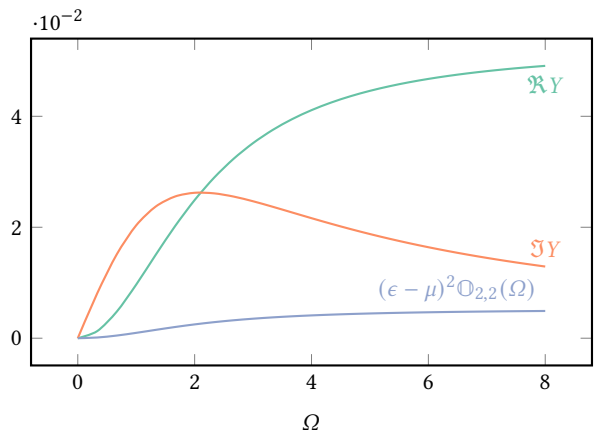

The dynamic response  in both frequency and time domain is computed plugging the above expressions into Equations \eqref{eq:fourier_me_solution} and \eqref{eq:prob_response_db_time} and rearranging them as 
\begin{align}
  \frac{\partial \hat{p}}{\partial \hat{f}_1}(\omega) &=- \frac{\epsilon j^{\mathrm{eq}}  }{ \left(w_{+1}^{\mathrm{eq}}+w_{-1}^{\mathrm{eq}}+i \omega \right)} = \epsilon \frac{\partial \hat{p}}{\partial \hat{f}_2} \label{eq:ex1_p_resp_f1}\,,
\end{align}
\begin{align}
      \frac{\delta p}{\delta f_1}(\tau) &=- \epsilon j^{\mathrm{eq}}  \e^{-(w_{+1}^{\mathrm{eq}}+w_{-1}^{\mathrm{eq}})\tau} \theta(\tau) =  \epsilon \frac{\delta p}{\delta f_2}(\tau) \,.\label{eq:ex_response_time}
\end{align}

We now specialize the analysis to the case of periodic driving on the temperature of the reservoir. The response function \eqref{eq:physcurrent_fields} of the currents  to a generic change of the entropic fields $\delta \hat{\vec{f}}$ is  
\begin{align}
     \nabla\hat{I}(\Omega) =
     \frac{\i \Omega j^{\mathrm{eq}}}{ w_{+1}^{\mathrm{eq}} + w_{-1}^{\mathrm{eq}}+\i \Omega }
\begin{pmatrix}
  \epsilon^2 & \epsilon\\
  \epsilon &  1
\end{pmatrix}\,.
\label{eq:sqd_phys_curr}
\end{align}
The generalized Onsager matrix  \eqref{eq:onsager_blocks} here consists of the $\mathbb{O}_\mathrm{pp}$ block only,
\begin{align}
     \mathbb{O}(\Omega)=\mathbb{O}_\mathrm{pp}(\Omega) = \frac{j^\mathrm{eq}\Omega^2 }{\left( (w_{+1}^{\mathrm{eq}}+ w_{-1}^{\mathrm{eq}})^2+ \Omega^2 \right)}
     \begin{pmatrix}
      \epsilon^2 & \epsilon\\
       \epsilon & 1
     \end{pmatrix}
     \label{eq:sqd_epr}
\end{align}
Since the system is tightly-coupled, both matrices have zero determinant, as it can be check directly multiplying on by the conservation law $\vec{\ell}_\mathrm{t.c.}$. Notice also that in Eqs.~\eqref{eq:sqd_phys_curr} and~\eqref{eq:sqd_epr} both the forward and backward equilibrium rates are needed, and this implies the full knowledge of the transition rates, not just their ratio as fixed by LDB, even if we are in presence of a single reservoir. When computing steady-state responses, however, this additional information on the transition rates is no longer needed, as it can be checked directly by imposing $\Omega=0$.

To make things more concrete, we compute the currents activated by a periodic perturbation on the temperature and the  dissipation associated to it. 
Note also that in terms of  $\beta$ and $\mu$ the probability displacement is
\begin{align}
    \delta p &= \left( \frac{\delta p}{\delta f_1} - \mu \frac{\delta p}{\delta f_2} \right) \delta \beta - \frac{\delta p}{\delta f_2} \delta \mu \\
             &= \frac{\delta p}{\delta f_2} \left( (\epsilon - \mu) \delta \beta -  \delta \mu \right)\,,
 \end{align}
 because of Eq.~\eqref{eq:ex1_p_resp_f1}. 
State observables are characterized by response functions that are directly obtained from the dynamic response, \textit{i.e.} $\delta \avg{ E} = \epsilon \delta p = \epsilon \delta \avg{N}$. The frequency-dependent response function of the energy to the temperature is then 
\begin{align}
\frac{\partial \avg{\hat{E}(\omega)}}{\partial \hat{\beta}} & = \epsilon (\epsilon - \mu) \frac{\partial \hat{p}}{\partial \hat{f}_2}(\omega) \,.
\end{align}
Note that its real and imaginary parts satisfy the Kramers--Kronig relations.
The Fourier representation of the current component $I$ in the periodic state for the perturbation $\delta \beta(t) = \e^{\i \Omega t} \delta \beta(0)$ is obtained from Equation \eqref{eq:physcurrent_fields} 
\begin{align}
  \delta \hat{\vec{I}}(\Omega)&= \avg{\mathbb{X}\tr, \mathbb{R}(\Omega)} \delta \hat{\vec{f}}(\Omega) 
\end{align}
and its response to temperature perturbation is obtained from the response to the entropic fields using the relation 
\begin{align}
    \delta \vec{I} = (\epsilon - \mu )  \frac{\partial I}{\partial f_2} \delta \beta - \beta \frac{\partial I}{\partial f_2} \delta \mu \,.
\end{align}
We can then introduce the thermal admittance defined as $Y(\Omega)\coloneq (\epsilon - \mu) \frac{\delta I}{\delta \hat{f}_2}$  with real and imaginary part given by 
\begin{align}
  \Re Y(\Omega)& = 
   (\epsilon -\mu) \frac{ \Omega ^2 j^\mathrm{eq} }{ \left(w_{-1}^{\mathrm{eq}} +w_{+1}^{\mathrm{eq}}\right)^2+\Omega ^2}\,, \label{eq:ry}\\
   \Im Y(\Omega) 
   &=  (\epsilon - \mu) 
   \frac{\Omega j^{\mathrm{eq}}(w_{-1}^{\mathrm{eq}} +w_{+1}^{\mathrm{eq}})}{
   (w_{-1}^{\mathrm{eq}} +w_{+1}^{\mathrm{eq}})^2 + \Omega^2} \,.\label{eq:iy}
\end{align}
The entropy production in this case is 
\begin{align}
    \overline{\dot\Sigma} \approx (\epsilon - \mu)^2 \mathbb{O}_{2,2}(\Omega) \delta \beta ^2\,.\label{eq:opp_sqd}
\end{align}
In Fig.~\ref{fig:stochastic_admittance} we show how Eqs.~\eqref{eq:ry}-\eqref{eq:opp_sqd} behave when varying the driving frequency.

\subsection{Electrical transport through a quantum dot}
\label{sec:ex_2}

We now reconsider Example~\ref{sec:ex_1} extending the setting with an additional reservoir of particles characterized by temperature and chemical potential $(\beta', \mu')$.

The incidence matrix after the addition of a second reservoir  becomes
\begin{align}
\mathbb{D} = \kbordermatrix{
  & +1 & +2  \cr
  0& -1& 1  \cr
  1& 1  & -1 } \,,
\end{align}
where a new transition path  labeled $+2$ appears. 
The vector of entropic fields is then given by
\begin{align}
  \vec{f} =
  \begin{pmatrix}
    \beta & -\beta \mu & \beta' & -\beta' \mu'
  \end{pmatrix}\tr\,.
\end{align}
The extensive quantities exchanged by the reservoirs with the system in transitions $+1$ and $+2$ are
\begin{align}
  \mathbb{X}\tr
  &=\kbordermatrix{
     & E_{\mathrm{l}}& N_{\mathrm{l}} & E_{\mathrm{r}} & N_{\mathrm{r}} \cr
    +1& \epsilon & 1 & 0 & 0\cr
    +2& 0        & 0 &-\epsilon & -1}\,.
\end{align}
Local detailed balance reads
\begin{align}
  \frac{w_{+1}}{w_{-1}} = \e^{-\beta (\epsilon - \mu)}\,,\qquad
  \frac{w_{+2}}{w_{-2}} = \e^{\beta' (\epsilon - \mu')}\,,
\end{align}
and the reference equilibrium  is defined setting $\beta =\beta'$ and $\mu = \mu'$. Under these conditions the rates satisfy detailed-balance for the equilibrium probability of occupancy $p^{\mathrm{eq}}=\e^{-\beta (\epsilon - \mu)}(1+\e^{-\beta (\epsilon - \mu)})^{-1}$.
The equilibrium fluxes are
\begin{align}
\vec{j}^{\mathrm{eq}}&=
  \begin{pmatrix}
    j_{+1}^{\mathrm{eq}} \\
   j_{+2}^{\mathrm{eq}}
  \end{pmatrix}
  =
  \begin{pmatrix}
   w_{+1}(1-p^{\mathrm{eq}}) \\
   w_{+2}p^{\mathrm{eq}}
  \end{pmatrix}\,.
\end{align}

Define $j^{\mathrm{eq}} \coloneq j^{\mathrm{eq}}_1+j^{\mathrm{eq}}_2$ and $w_\pm \coloneq w_{\pm 1} + w_{\pm 2}$. The matrices that appear in the response functions are then
\begin{align}
  \i \omega \mathbb{P} + \avg{\mathbb{D}\tr,\mathbb{D}\tr}
  &=
\begin{pmatrix}
          \i \omega (1 - p^{\mathrm{eq}}) + j^{\mathrm{eq}}   & -  j^{\mathrm{eq}} \\
       -  j^{\mathrm{eq}} &  \i \omega p^{\mathrm{eq}} \\
\end{pmatrix}
  \\
  \avg{\mathbb{D}\tr, \mathbb{X}\tr}
  &=  j^{\mathrm{eq}}
      \begin{pmatrix}
         -\epsilon  & -1 & - \epsilon & - 1\\
       \epsilon  & 1 & \epsilon & 1 \\
      \end{pmatrix}
\end{align}
The dynamic response of the occupation probability to a generic perturbation $\delta \vec{f}$ is
\begin{align}
\delta\hat{p}(\omega) = -\frac{1}{2} \frac{w^{\mathrm{eq}}_+}{(w^{\mathrm{eq}}_+ + w^{\mathrm{eq}}_- )\ + \i \omega }
      \begin{pmatrix}
        \epsilon \,,  & 1\,, &\epsilon \,, & 1 \\
     \end{pmatrix}\tr \delta \hat{\vec{f}}(\omega)\,.
\end{align}
This system has a single cycle $\vec{C}=(+1,-1)$ in $\ker \mathbb{D}$. thus $\mathbb{M}=\mathbb{X}\mathbb{C}=(\epsilon, 1, -\epsilon,-1)$. The conservation laws are thus given by the equation $\epsilon x_{1} +x_{2} -\epsilon x_{3}-x_{4}=0$. Notice that the system is tight-coupled, as in addition to the one representing the conservation of charge and particles $\vec{\ell}_{E}=(1,0,1,0)$ and $\vec{\ell}_{N}=(0,1,0,1)$ we have the proportionality link between the exchanged charges and particles $\vec{\ell}_{\mathrm{t.c.}}=(0,0,1,-\epsilon)$.
The only resulting nonconservative force is
\begin{align}
    \delta \mathcal{F} =  -(\mu'-\epsilon)\beta'- (\mu-\epsilon) \beta\,,\label{eq:sqd2_force}
\end{align}
which can be read as an effective difference of chemical potentials rescaled by inverse temperatures.
Notice how this nonconservative force depends explicitly on the system quantity $\epsilon$, and hence on the protocol.
However, notice also that this dependence disappears when $\beta=\beta'$. 

We now consider a perturbation with a single frequency $\Omega$.
The physical currents  and the entropy production response matrices read respectively
\begin{widetext}
\begin{align}
  \mathbb{T}\tr\nabla \hat{\vec{I}}(\Omega) \mathbb{T} =\frac{1}{ \left(w_{+}^{\mathrm{eq}} +w_{-}^{\mathrm{eq}}+\i \Omega \right)}
\begin{pmatrix}
 j^{\mathrm{eq}}_{+1} \epsilon ^2  \left(w_{+2}^{\mathrm{eq}}+w_{-2}^{\mathrm{eq}} +\i \Omega \right) & 0 & -\i j_{+1}^{\mathrm{eq}} \Omega  \epsilon ^2  & -\i j_{+1}^{\mathrm{eq}} \Omega  \epsilon   \\
 0 & 0 & 0 & 0 \\
 -\i j_{+1}^{\mathrm{eq}} \Omega  \epsilon ^2  & 0 & \i \Omega  \epsilon ^2 (j_{+1}^{\mathrm{eq}}+j_{+2}^{\mathrm{eq}})  & \i \Omega  \epsilon  (j_{+1}^{\mathrm{eq}}+j_{+2}^{\mathrm{eq}}) \\
 -\i j_{+1}^{\mathrm{eq}} \Omega  \epsilon   & 0 & \i \Omega  \epsilon  (j_{+1}^{\mathrm{eq}}+j_{+2}^{\mathrm{eq}})  & \i \Omega  (j_{+1}^{\mathrm{eq}}+j_{+2}^{\mathrm{eq}})
\end{pmatrix}\,,
\label{eq:curr_sqd2}
\end{align}
\begin{align}
    \mathbb{O}(\Omega)=
    \frac{1}{\left[ (w_{+}^\mathrm{eq} + w_{-}^{\mathrm{eq}})^2+\Omega ^2\right]}
    \begin{pmatrix}
 j_{+1}^{\mathrm{eq}} \epsilon ^2  \left[ \left(w_{+}^{\mathrm{eq}} + w_{-}^{\mathrm{eq}} \right) \left(w_{+2}^{\mathrm{eq}} +w_{-2}^{\mathrm{eq}}  \right) +\Omega ^2 \right] & 0 & -j_{+1}^{\mathrm{eq}} \Omega ^2 \epsilon ^2 &- j_{+1}^{\mathrm{eq}} \Omega ^2 \epsilon  \\
 0 & 0 & 0 & 0 \\
-j_{+1}^{\mathrm{eq}} \Omega ^2 \epsilon ^2   & 0 & \Omega ^2 \epsilon ^2 j_{+}^{\mathrm{eq}}& \Omega ^2 \epsilon  j_{+}^{\mathrm{eq}}  \\
- j_{+}^{\mathrm{eq}} \Omega ^2 \epsilon  & 0 & \Omega ^2 \epsilon  j_{+}^{\mathrm{eq}}& \Omega ^2 j_{+}^{\mathrm{eq}}
    \end{pmatrix}\,.
    \label{eq:epr_sqd2}
\end{align}
\end{widetext}
As discussed in \S\ref{sec:currents_general}, the existence of a tight coupling conservation law makes these two matrices degenerate at every frequency.
Notice, furthermore, that these two matrices coincide at steady state, \textit{i.e.} when $\Omega=0$, and that the LDB condition becomes sufficient to obtain the two response matrices, instead of the full knowledge of the transition rates for forward and backward transitions.

\subsection{Photoelectric two-levels nanodevice}
\label{sec:ex_3} 

\begin{figure}
      \subfloat[Schematic representation]{
      \def\svgwidth{0.3\textwidth}
      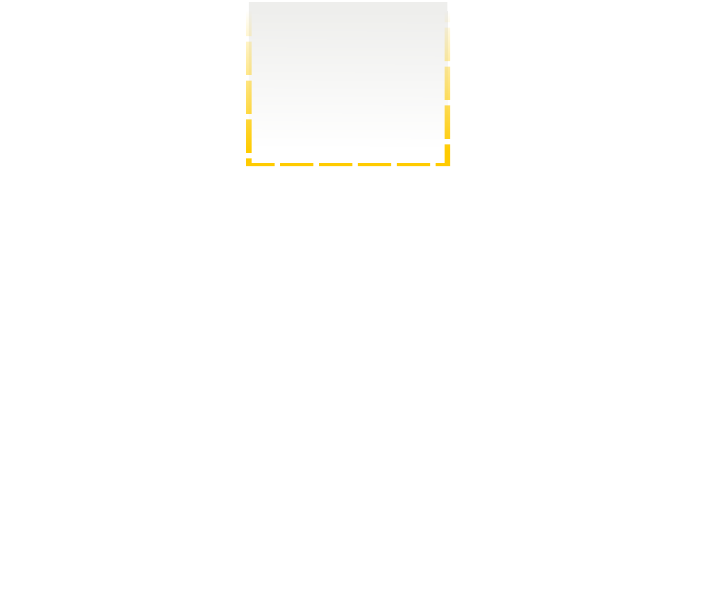}
    \quad
    \subfloat[Network]{
      \def\svgwidth{0.18\textwidth}
      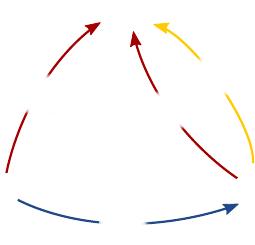}
    \quad
     \subfloat[Cycles]{
      \def\svgwidth{0.36\textwidth}
      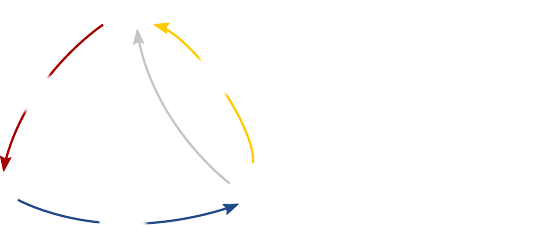}
    \caption{
      Two-levels nanodevice powered by incoming radiation. (a) Schematics: the nanodevice is connected to two reservoirs of particles, at different chemical potentials. The energy from the incoming radiation allows for transport against this gradient. (b) Network of transitions between microscopic states. (c) Decomposition into fundamental cycles.}
    \label{fig:solarcell}
\end{figure}

As a final example, we consider a photoelectric nanodevice that uses energy from incoming radiation to transport particles between two reservoirs maintained at different chemical potential \cite{rutten2009reaching}, depicted in Fig.~\ref{fig:solarcell}.
In this model the transport of energy and the one of matter are decoupled, as there exists a transition between the two levels that is mediated by the absorption or emission of a photon, and not by an exchange of electrons with the reservoirs. This implies that the system is not tight-coupled for generic choices of the parameters, as we will show in the following.

The incidence matrix reads
\begin{align}
\mathbb{D} = \kbordermatrix{
  & +1 & +2  & +3 & +4\cr
  0& -1& -1&  0&  0  \cr
  1&  1&  0& -1& -1 \cr
  2&  0&  1&  1& 1} \, ,
\end{align}
where the different labels refer to the transitions depicted in Fig.~\ref{fig:solarcell}.
The vector of entropic fields is then given by
\begin{align}
  \vec{f} =
  \begin{pmatrix}
    \beta & -\beta \mu_1 & -\beta \mu_2 & \beta_\mathrm{s}
  \end{pmatrix}\tr\,.
\end{align}
The extensive quantities exchanged by the reservoirs with the system in transitions $+1$ and $+2$ are
\begin{align}
  \mathbb{X}\tr
  &=\kbordermatrix{
     & E_{\mathrm{g}}& N_{\mathrm{l}} &  N_{\mathrm{r}} &E_{\mathrm{s}} \cr
    +1& \epsilon_\mathrm{d} & 1 & 0 & 0\cr
    +2& \epsilon_\mathrm{u} & 0 &1 & 0\cr
    +3& 0 & 0& 0& \epsilon_{\mathrm{u}}-\epsilon_{\mathrm{d}} \cr
    +4& \epsilon_{\mathrm{u}}-\epsilon_{\mathrm{d}} &0 &0 &0}\,.
\end{align}
and the reference equilibrium  is achieved when the temperature of the radiative source is the same as the one of the ground, $\beta_\mathrm{s}=\beta$, and when the two reservoirs have the same chemical potential, $\mu_2=\mu_1$.

The conservation law vectors are $\vec{\ell}_1=(1,0,0,1)$ and $\vec{\ell}_2=(0,1,1,0)$, and they are related through the balance \eqref{eq:conservationlaws} respectively to the conservation of energy and particles. When $\beta, \mu_1$ are considered as fixing the reference equilibrium, the corresponding fundamental forces are $\vec{\mathcal{F}}_1=\beta(\mu_2-\mu_1) $ and  $\vec{\mathcal{F}}_2=\beta-\beta_\mathrm{s}$.

For this model, the generalized Onsager matrix $\mathbb{O}(\omega)$ and response matrix of the current $\mathbb{Q}$ can be obtained analytically, using the same procedure of the previous examples. However, the  resulting expressions of the two matrices are not compact enough to be included. Nevertheless, we can still make some conceptually important observation by studying the determinant of the generalized Onsager matrix evaluated in the limit $\omega\to 0$, that is 
\begin{align}
    \det \mathbb{O}(0) = 16
    \frac{ (\epsilon \mathrm{d}-\epsilon_\mathrm{u})^4 \e^{\beta  (3 \mu_1 +\mu_2+2 \epsilon_\mathrm{d}+2 \epsilon_\mathrm{u})+\beta_s (\epsilon_\mathrm{d}-\epsilon_\mathrm{u})}}{\left(\e^{\beta  (\mu_1+\epsilon_\mathrm{d})}+\e^{\beta  (\epsilon_\mathrm{d}+\epsilon_\mathrm{u})}+\e^{\beta  (\mu_1+\epsilon_u)}\right)^4}\,.
\end{align}
Since it is proportional to a power of the difference between the energy levels $(\epsilon_d - \epsilon_u)$, the matrix becomes singular when the two levels have the same energy. This condition is the same for which the system becomes tight-coupled. In fact, if $\epsilon_\mathrm{d}=\epsilon_\mathrm{u}$, a new conservation law appears, namely $\vec{\ell}_\mathrm{t.c}=(0,0,0,1)$, that is in the cokernel of $\mathbb{X}$.

\section{Conclusions and outlook}
\label{sec:conclusions}

We now summarize our results going from the most general to the special cases covered by our theory.

The near equilibrium decomposition of the EPR in Eq.~\eqref{eq:sstransient} is a direct consequence of the systematic introduction of conservation laws in stochastic thermodynamics ~\cite{rao2018conservation}. This decomposition describes the dissipation at finite times caused by a general, non-periodic protocol acting both on the extensive quantities of the states of the system and on the intensive parameters of the reservoirs. The decomposition strongly constrains the dynamics near equilibrium, as it immediately implies the minimum entropy production principle, when only steady thermodynamic forces are considered.
In our formulation, the EPR is made up of two separately positive contributions. The first one is a quadratic form in the vector $\delta \vec{\mathcal{F}}$ of fundamental thermodynamic forces and identifies the static part of the Onsager matrix. This term accounts for the dissipation due to maintaining a steady state, while the other contribution describes the dissipation due to purely dynamical effects as it vanishes at steady state or for adiabatic driving.

In the special case of periodic driving, the system settles on a periodic steady state. We derived the response matrix~\eqref{eq:currents_general} for the Fourier components of the currents, concluding that in the linear regime it is not possible to generate currents against a nonconservative force by only performing time-dependent driving on the fields that define the equilibrium. Response coefficients for conserved quantities can also be derived and expressed via Green--Kubo-like formulas \eqref{eq:gk_most_general}, and this general expression reduces to the classic one~\eqref{eq:gk} when detailed balanced protocols are considered. The EPR \eqref{eq:epr_most_general} allows to identify the minimal subset of generalized thermodynamic forces that define a periodic protocol, namely the nonconservative forces \eqref{eq:fundForce} and the Massieu potential \eqref{eq:potential} of each state of the system.  
The decomposition~\eqref{eq:epr_most_general}, as well as the more general~\eqref{eq:sstransient}, also describes the regime of adiabatic driving, for which the only dissipation and currents are associated to the existence of nonzero average nonconservative forces.

Finally, restricting ourselves to protocols that only act on the reservoir intensive parameters, we obtained an expression for
the response of the currents~\eqref{eq:physicalcurr_response} and for the EPR in Eq.~\eqref{eq:onsagerdef}, and we call the latter a generalized Onsager matrix (for each frequency mode present in the driving).  To account for all possible mechanisms of dissipation, the set of fundamental forces must be extended by using the potential fields that define the time-dependent equilibrium. Employing the formalism of conservation laws~\cite{rao2018conservation}, we are able to construct these matrices such that they are always symmetric. Even if the explicit time dependence of the protocol does not destroy the symmetry,  it makes the response matrix of the currents different from the generalized Onsager matrix~\eqref{eq:onsagerdef}, unless steady perturbations are used.  Our conclusion is that time-dependent driving near equilibrium does not break the time-reversal symmetry of the dynamics as far as response coefficients are concerned, as these are computed employing equilibrium quantities only. This provides an alternative point of view compared to that presented in  Refs.~\cite{brandner2016periodic, proesmans2019general}, where a single, non-symmetric and protocol-dependent matrix was used to obtain both the time averaged currents and the EPR. Furthermore, we proved the connection between the existence of tight-coupling conservation laws and the vanishing of the determinant of the generalized Onsager matrix.

The present work always addresses the thermodynamics of linear response around equilibrium. Linear response of far-from-equilibrium steady steady has been explored using stochastic thermodynamics in recent years~\cite{speck2009extended, baiesi2013response, altaner2016fluctuation, horowitz2019thermodynamic} but the implications of a proper identification of the thermodynamics forces in such situations is left for future work. 
This work also provides a useful starting point to design optimal protocols in the linear regime. In fact, Eqs.~\eqref{eq:epr_bilinear2} and~\eqref{eq:epronsager} allow for straightforward optimization procedures to obtain protocols that minimize dissipation in steady or periodic working conditions, at least in the linear regime. This complements previous studies on optimal protocols in the linear regime \cite{bauer2016optimal, proesmans2016power} as well as the approaches based on thermodynamic length~\cite{crooks2007measuring, sivak2012thermodynamic, zulkowski2012geometry, mandal2016analysis}, that are concerned with transient transformations between different steady states.

This work was supported by grants from the Simons Foundation (691552, RR), and the European Research Council, project NanoThermo (ERC-2015-CoG Agreement No.~681456, DF and ME).

%%% Local Variables:
%%% mode: latex
%%% TeX-master: "../article"
%%% End:

%% file: figures/systemBaths.pdf_tex
%% Creator: Inkscape 1.1.1 (3bf5ae0d25, 2021-09-20, custom), www.inkscape.org
%% Accompanies image file 'figures/systemBaths.pdf' (pdf, eps, ps)
\begingroup%
  \makeatletter%
  \providecommand\color[2][]{%
    \errmessage{(Inkscape) Color is used for the text in Inkscape, but the package 'color.sty' is not loaded}%
    \renewcommand\color[2][]{}%
  }%
  \providecommand\transparent[1]{%
    \errmessage{(Inkscape) Transparency is used (non-zero) for the text in Inkscape, but the package 'transparent.sty' is not loaded}%
    \renewcommand\transparent[1]{}%
  }%
  \providecommand\rotatebox[2]{#2}%
  \newcommand*\fsize{\dimexpr\f@size pt\relax}%
  \newcommand*\lineheight[1]{\fontsize{\fsize}{#1\fsize}\selectfont}%
  \ifx\svgwidth\undefined%
    \setlength{\unitlength}{268.79419938bp}%
    \ifx\svgscale\undefined%
      \relax%
    \else%
      \setlength{\unitlength}{\unitlength * \real{\svgscale}}%
    \fi%
  \else%
    \setlength{\unitlength}{\svgwidth}%
  \fi%
  \global\let\svgwidth\undefined%
  \global\let\svgscale\undefined%
  \makeatother%
  \begin{picture}(1,0.52750735)%
    \lineheight{1}%
    \setlength\tabcolsep{0pt}%
    \put(0.02461797,0.34380708){\color[rgb]{0,0,0}\makebox(0,0)[lt]{\begin{minipage}{1.05648761\unitlength}\raggedright  \end{minipage}}}%
    \put(0,0){\includegraphics[width=\unitlength,page=1]{figures/systemBaths.pdf}}%
    \put(0.44290781,0.51406905){\color[rgb]{0,0,0}\makebox(0,0)[lt]{\lineheight{0}\smash{\begin{tabular}[t]{l}physical\\reservoirs $\{\rho\}$\end{tabular}}}}%
    \put(0.68412798,0.12328692){\color[rgb]{0,0,0}\makebox(0,0)[lt]{\lineheight{0}\smash{\begin{tabular}[t]{l}edges $\{e\}$\end{tabular}}}}%
    \put(0.44289648,0.03001565){\color[rgb]{0,0,0}\makebox(0,0)[lt]{\lineheight{0}\smash{\begin{tabular}[t]{l}generalized\\reservoirs $\{y\}=\{(\rho,\kappa)\}$\end{tabular}}}}%
    \put(0.31728452,0.12375155){\color[rgb]{0,0,0}\makebox(0,0)[lt]{\lineheight{0}\smash{\begin{tabular}[t]{l}states $\{n\}$\end{tabular}}}}%
    \put(0.49761338,0.16188886){\color[rgb]{0,0,0}\makebox(0,0)[lt]{\lineheight{0}\smash{\begin{tabular}[t]{l}system\end{tabular}}}}%
    \put(0.20386679,0.30773305){\color[rgb]{0,0,0}\makebox(0,0)[lt]{\lineheight{0}\smash{\begin{tabular}[t]{l}$\Delta N \rightarrow -\mu_1\beta_1$\end{tabular}}}}%
    \put(0.20944727,0.33463551){\color[rgb]{0,0,0}\makebox(0,0)[lt]{\lineheight{0}\smash{\begin{tabular}[t]{l} $\Delta E \rightarrow \beta_1 $\end{tabular}}}}%
    \put(0.69923464,0.30514832){\color[rgb]{0,0,0}\makebox(0,0)[lt]{\lineheight{0}\smash{\begin{tabular}[t]{l} $\Delta N \rightarrow -\mu_2\beta_2$\end{tabular}}}}%
    \put(0.70447695,0.3372204){\color[rgb]{0,0,0}\makebox(0,0)[lt]{\lineheight{0}\smash{\begin{tabular}[t]{l}$\Delta E \rightarrow \beta_2 $\end{tabular}}}}%
    \put(0.54540839,-0.01885132){\color[rgb]{0,0,0}\makebox(0,0)[lt]{\begin{minipage}{0.58601584\unitlength}\end{minipage}}}%
    \put(-0.00061767,0.38636525){\color[rgb]{0,0,0}\makebox(0,0)[lt]{\lineheight{1.25}\smash{\begin{tabular}[t]{l}Exchanged quantities $\mathbb{Y}_{\kappa}$\end{tabular}}}}%
    \put(0.72447483,0.38425997){\color[rgb]{0,0,0}\makebox(0,0)[lt]{\lineheight{1.25}\smash{\begin{tabular}[t]{l}Entropic fields $\vec{f}$\end{tabular}}}}%
  \end{picture}%
\endgroup%

%% file: figures/singlebathqd.pdf_tex
%% Creator: Inkscape inkscape 0.92.5, www.inkscape.org
%% Accompanies image file 'singlebathqd.pdf' (pdf, eps, ps)
\begingroup%
  \makeatletter%
  \providecommand\color[2][]{%
    \errmessage{(Inkscape) Color is used for the text in Inkscape, but the package 'color.sty' is not loaded}%
    \renewcommand\color[2][]{}%
  }%
  \providecommand\transparent[1]{%
    \errmessage{(Inkscape) Transparency is used (non-zero) for the text in Inkscape, but the package 'transparent.sty' is not loaded}%
    \renewcommand\transparent[1]{}%
  }%
  \providecommand\rotatebox[2]{#2}%
  \newcommand*\fsize{\dimexpr\f@size pt\relax}%
  \newcommand*\lineheight[1]{\fontsize{\fsize}{#1\fsize}\selectfont}%
  \ifx\svgwidth\undefined%
    \setlength{\unitlength}{107.8480686bp}%
    \ifx\svgscale\undefined%
      \relax%
    \else%
      \setlength{\unitlength}{\unitlength * \real{\svgscale}}%
    \fi%
  \else%
    \setlength{\unitlength}{\svgwidth}%
  \fi%
  \global\let\svgwidth\undefined%
  \global\let\svgscale\undefined%
  \makeatother%
  \begin{picture}(1,0.54504741)%
    \lineheight{1}%
    \setlength\tabcolsep{0pt}%
    \put(0,0){\includegraphics[width=\unitlength,page=1]{singlebathqd.pdf}}%
    \put(0.64572308,0.33212113){\color[rgb]{0,0,0}\makebox(0,0)[lt]{\lineheight{0}\smash{\begin{tabular}[t]{l}$\beta$\end{tabular}}}}%
    \put(0.64572308,0.1718246){\color[rgb]{0,0,0}\makebox(0,0)[lt]{\lineheight{0}\smash{\begin{tabular}[t]{l}$\mu$\end{tabular}}}}%
    \put(0.15261299,0.0341112){\color[rgb]{0,0,0}\makebox(0,0)[lt]{\lineheight{0}\smash{\begin{tabular}[t]{l}QD\end{tabular}}}}%
    \put(0,0){\includegraphics[width=\unitlength,page=2]{singlebathqd.pdf}}%
  \end{picture}%
\endgroup%

%% file: figures/networksbqd.pdf_tex
%% Creator: Inkscape inkscape 0.92.5, www.inkscape.org
%% Accompanies image file '"figures/networksbqd.pdf"' (pdf, eps, ps)
\begingroup%
  \makeatletter%
  \providecommand\color[2][]{%
    \errmessage{(Inkscape) Color is used for the text in Inkscape, but the package 'color.sty' is not loaded}%
    \renewcommand\color[2][]{}%
  }%
  \providecommand\transparent[1]{%
    \errmessage{(Inkscape) Transparency is used (non-zero) for the text in Inkscape, but the package 'transparent.sty' is not loaded}%
    \renewcommand\transparent[1]{}%
  }%
  \providecommand\rotatebox[2]{#2}%
  \newcommand*\fsize{\dimexpr\f@size pt\relax}%
  \newcommand*\lineheight[1]{\fontsize{\fsize}{#1\fsize}\selectfont}%
  \ifx\svgwidth\undefined%
    \setlength{\unitlength}{79.61285732bp}%
    \ifx\svgscale\undefined%
      \relax%
    \else%
      \setlength{\unitlength}{\unitlength * \real{\svgscale}}%
    \fi%
  \else%
    \setlength{\unitlength}{\svgwidth}%
  \fi%
  \global\let\svgwidth\undefined%
  \global\let\svgscale\undefined%
  \makeatother%
  \begin{picture}(1,0.83441885)%
    \lineheight{1}%
    \setlength\tabcolsep{0pt}%
    \put(-0.00489871,0.16695628){\color[rgb]{0,0,0}\makebox(0,0)[lt]{\lineheight{0}\smash{\begin{tabular}[t]{l}0\end{tabular}}}}%
    \put(0,0){\includegraphics[width=\unitlength,page=1]{"figures/networksbqd.pdf"}}%
    \put(0.89132496,0.16695628){\color[rgb]{0,0,0}\makebox(0,0)[lt]{\lineheight{0}\smash{\begin{tabular}[t]{l}1\end{tabular}}}}%
    \put(0,0){\includegraphics[width=\unitlength,page=2]{"figures/networksbqd.pdf"}}%
    \put(0.38634356,0.07622689){\color[rgb]{0.64313725,0,0}\makebox(0,0)[lt]{\lineheight{0}\smash{\begin{tabular}[t]{l}\scriptsize+1\end{tabular}}}}%
  \end{picture}%
\endgroup%

%% file: figures/transportqd.pdf_tex
%% Creator: Inkscape inkscape 0.92.5, www.inkscape.org
%% Accompanies image file 'transportqd.pdf' (pdf, eps, ps)
\begingroup%
  \makeatletter%
  \providecommand\color[2][]{%
    \errmessage{(Inkscape) Color is used for the text in Inkscape, but the package 'color.sty' is not loaded}%
    \renewcommand\color[2][]{}%
  }%
  \providecommand\transparent[1]{%
    \errmessage{(Inkscape) Transparency is used (non-zero) for the text in Inkscape, but the package 'transparent.sty' is not loaded}%
    \renewcommand\transparent[1]{}%
  }%
  \providecommand\rotatebox[2]{#2}%
  \newcommand*\fsize{\dimexpr\f@size pt\relax}%
  \newcommand*\lineheight[1]{\fontsize{\fsize}{#1\fsize}\selectfont}%
  \ifx\svgwidth\undefined%
    \setlength{\unitlength}{202.80028576bp}%
    \ifx\svgscale\undefined%
      \relax%
    \else%
      \setlength{\unitlength}{\unitlength * \real{\svgscale}}%
    \fi%
  \else%
    \setlength{\unitlength}{\svgwidth}%
  \fi%
  \global\let\svgwidth\undefined%
  \global\let\svgscale\undefined%
  \makeatother%
  \begin{picture}(1,0.2898532)%
    \lineheight{1}%
    \setlength\tabcolsep{0pt}%
    \put(0,0){\includegraphics[width=\unitlength,page=1]{transportqd.pdf}}%
    \put(0.12259102,0.17662018){\color[rgb]{0,0,0}\makebox(0,0)[lt]{\lineheight{0}\smash{\begin{tabular}[t]{l}$\beta$\end{tabular}}}}%
    \put(0.12259102,0.09137537){\color[rgb]{0,0,0}\makebox(0,0)[lt]{\lineheight{0}\smash{\begin{tabular}[t]{l}$\mu$\end{tabular}}}}%
    \put(0,0){\includegraphics[width=\unitlength,page=2]{transportqd.pdf}}%
    \put(0.81159745,0.17662018){\color[rgb]{0,0,0}\makebox(0,0)[lt]{\lineheight{0}\smash{\begin{tabular}[t]{l}$\beta'$\end{tabular}}}}%
    \put(0.81159745,0.09137537){\color[rgb]{0,0,0}\makebox(0,0)[lt]{\lineheight{0}\smash{\begin{tabular}[t]{l}$\mu'$\end{tabular}}}}%
    \put(0.4613156,-0.00867749){\color[rgb]{0,0,0}\makebox(0,0)[lt]{\lineheight{0}\smash{\begin{tabular}[t]{l}QD\end{tabular}}}}%
    \put(0,0){\includegraphics[width=\unitlength,page=3]{transportqd.pdf}}%
  \end{picture}%
\endgroup%

%% file: figures/network2bqd.pdf_tex
%% Creator: Inkscape 1.0.2 (e86c870879, 2021-01-15), www.inkscape.org
%% Accompanies image file 'figures/network2bqd.pdf' (pdf, eps, ps)
\begingroup%
  \makeatletter%
  \providecommand\color[2][]{%
    \errmessage{(Inkscape) Color is used for the text in Inkscape, but the package 'color.sty' is not loaded}%
    \renewcommand\color[2][]{}%
  }%
  \providecommand\transparent[1]{%
    \errmessage{(Inkscape) Transparency is used (non-zero) for the text in Inkscape, but the package 'transparent.sty' is not loaded}%
    \renewcommand\transparent[1]{}%
  }%
  \providecommand\rotatebox[2]{#2}%
  \newcommand*\fsize{\dimexpr\f@size pt\relax}%
  \newcommand*\lineheight[1]{\fontsize{\fsize}{#1\fsize}\selectfont}%
  \ifx\svgwidth\undefined%
    \setlength{\unitlength}{79.5749636bp}%
    \ifx\svgscale\undefined%
      \relax%
    \else%
      \setlength{\unitlength}{\unitlength * \real{\svgscale}}%
    \fi%
  \else%
    \setlength{\unitlength}{\svgwidth}%
  \fi%
  \global\let\svgwidth\undefined%
  \global\let\svgscale\undefined%
  \makeatother%
  \begin{picture}(1,1.29591072)%
    \lineheight{1}%
    \setlength\tabcolsep{0pt}%
    \put(-0.00394246,0.16808087){\color[rgb]{0,0,0}\makebox(0,0)[lt]{\lineheight{0}\smash{\begin{tabular}[t]{l}0\end{tabular}}}}%
    \put(0,0){\includegraphics[width=\unitlength,page=1]{figures/network2bqd.pdf}}%
    \put(0.87403006,0.1678328){\color[rgb]{0,0,0}\makebox(0,0)[lt]{\lineheight{0}\smash{\begin{tabular}[t]{l}1\end{tabular}}}}%
    \put(0,0){\includegraphics[width=\unitlength,page=2]{figures/network2bqd.pdf}}%
    \put(0.37658468,0.31663697){\color[rgb]{0.1254902,0.29019608,0.52941176}\makebox(0,0)[lt]{\lineheight{0}\smash{\begin{tabular}[t]{l}\scriptsize+2\end{tabular}}}}%
    \put(0,0){\includegraphics[width=\unitlength,page=3]{figures/network2bqd.pdf}}%
    \put(0.38605135,0.03571004){\color[rgb]{0.64313725,0,0}\makebox(0,0)[lt]{\lineheight{0}\smash{\begin{tabular}[t]{l}\scriptsize+1\end{tabular}}}}%
  \end{picture}%
\endgroup%

%% file: figures/solarcell.pdf_tex
%% Creator: Inkscape 1.1.2 (0a00cf5339, 2022-02-04, custom), www.inkscape.org
%% Accompanies image file 'figures/solarcell.pdf' (pdf, eps, ps)
\begingroup%
  \makeatletter%
  \providecommand\color[2][]{%
    \errmessage{(Inkscape) Color is used for the text in Inkscape, but the package 'color.sty' is not loaded}%
    \renewcommand\color[2][]{}%
  }%
  \providecommand\transparent[1]{%
    \errmessage{(Inkscape) Transparency is used (non-zero) for the text in Inkscape, but the package 'transparent.sty' is not loaded}%
    \renewcommand\transparent[1]{}%
  }%
  \providecommand\rotatebox[2]{#2}%
  \newcommand*\fsize{\dimexpr\f@size pt\relax}%
  \newcommand*\lineheight[1]{\fontsize{\fsize}{#1\fsize}\selectfont}%
  \ifx\svgwidth\undefined%
    \setlength{\unitlength}{204.8823278bp}%
    \ifx\svgscale\undefined%
      \relax%
    \else%
      \setlength{\unitlength}{\unitlength * \real{\svgscale}}%
    \fi%
  \else%
    \setlength{\unitlength}{\svgwidth}%
  \fi%
  \global\let\svgwidth\undefined%
  \global\let\svgscale\undefined%
  \makeatother%
  \begin{picture}(1,0.86484216)%
    \lineheight{1}%
    \setlength\tabcolsep{0pt}%
    \put(0,0){\includegraphics[width=\unitlength,page=1]{figures/solarcell.pdf}}%
    \put(0.45006289,0.69347007){\color[rgb]{0,0,0}\makebox(0,0)[lt]{\lineheight{0}\smash{\begin{tabular}[t]{l}$\beta_\text{s}$\end{tabular}}}}%
    \put(0,0){\includegraphics[width=\unitlength,page=2]{figures/solarcell.pdf}}%
    \put(0.05545436,0.17525134){\color[rgb]{0,0,0}\makebox(0,0)[lt]{\lineheight{0}\smash{\begin{tabular}[t]{l}$\beta$\end{tabular}}}}%
    \put(0.05545436,0.0908728){\color[rgb]{0,0,0}\makebox(0,0)[lt]{\lineheight{0}\smash{\begin{tabular}[t]{l}$\mu_1$\end{tabular}}}}%
    \put(0,0){\includegraphics[width=\unitlength,page=3]{figures/solarcell.pdf}}%
    \put(0.8033505,0.17525134){\color[rgb]{0,0,0}\makebox(0,0)[lt]{\lineheight{0}\smash{\begin{tabular}[t]{l}$\beta$\end{tabular}}}}%
    \put(0.8033505,0.0908728){\color[rgb]{0,0,0}\makebox(0,0)[lt]{\lineheight{0}\smash{\begin{tabular}[t]{l}$\mu_2$\end{tabular}}}}%
    \put(0,0){\includegraphics[width=\unitlength,page=4]{figures/solarcell.pdf}}%
    \put(0.6824515,0.41593412){\color[rgb]{0,0,0}\makebox(0,0)[lt]{\lineheight{0}\smash{\begin{tabular}[t]{l}u\end{tabular}}}}%
    \put(0,0){\includegraphics[width=\unitlength,page=5]{figures/solarcell.pdf}}%
    \put(0.42760674,0.18677705){\color[rgb]{0,0,0}\makebox(0,0)[lt]{\lineheight{0}\smash{\begin{tabular}[t]{l}d\end{tabular}}}}%
    \put(0,0){\includegraphics[width=\unitlength,page=6]{figures/solarcell.pdf}}%
  \end{picture}%
\endgroup%

%% file: figures/network_solar.pdf_tex
%% Creator: Inkscape inkscape 0.92.5, www.inkscape.org
%% Accompanies image file 'network_solar.pdf' (pdf, eps, ps)
\begingroup%
  \makeatletter%
  \providecommand\color[2][]{%
    \errmessage{(Inkscape) Color is used for the text in Inkscape, but the package 'color.sty' is not loaded}%
    \renewcommand\color[2][]{}%
  }%
  \providecommand\transparent[1]{%
    \errmessage{(Inkscape) Transparency is used (non-zero) for the text in Inkscape, but the package 'transparent.sty' is not loaded}%
    \renewcommand\transparent[1]{}%
  }%
  \providecommand\rotatebox[2]{#2}%
  \newcommand*\fsize{\dimexpr\f@size pt\relax}%
  \newcommand*\lineheight[1]{\fontsize{\fsize}{#1\fsize}\selectfont}%
  \ifx\svgwidth\undefined%
    \setlength{\unitlength}{73.23962655bp}%
    \ifx\svgscale\undefined%
      \relax%
    \else%
      \setlength{\unitlength}{\unitlength * \real{\svgscale}}%
    \fi%
  \else%
    \setlength{\unitlength}{\svgwidth}%
  \fi%
  \global\let\svgwidth\undefined%
  \global\let\svgscale\undefined%
  \makeatother%
  \begin{picture}(1,0.9631027)%
    \lineheight{1}%
    \setlength\tabcolsep{0pt}%
    \put(-0.005325,0.15662177){\color[rgb]{0,0,0}\makebox(0,0)[lt]{\lineheight{0}\smash{\begin{tabular}[t]{l}0\end{tabular}}}}%
    \put(0.94859334,0.15635224){\color[rgb]{0,0,0}\makebox(0,0)[lt]{\lineheight{0}\smash{\begin{tabular}[t]{l}1\end{tabular}}}}%
    \put(0.463809,0.89734587){\color[rgb]{0,0,0}\makebox(0,0)[lt]{\lineheight{0}\smash{\begin{tabular}[t]{l}2\end{tabular}}}}%
    \put(0,0){\includegraphics[width=\unitlength,page=1]{network_solar.pdf}}%
    \put(0.40659308,0.05933145){\color[rgb]{0.1254902,0.29019608,0.52941176}\makebox(0,0)[lt]{\lineheight{0}\smash{\begin{tabular}[t]{l}+1\end{tabular}}}}%
    \put(0.7893772,0.62715746){\color[rgb]{1,0.79607843,0}\makebox(0,0)[lt]{\lineheight{0}\smash{\begin{tabular}[t]{l}+3\end{tabular}}}}%
    \put(0.08172054,0.57376397){\color[rgb]{0.64313725,0,0}\makebox(0,0)[lt]{\lineheight{0}\smash{\begin{tabular}[t]{l}+2\end{tabular}}}}%
    \put(0.60578059,0.49425537){\color[rgb]{0.64313725,0,0}\makebox(0,0)[lt]{\lineheight{0}\smash{\begin{tabular}[t]{l}+4\end{tabular}}}}%
  \end{picture}%
\endgroup%

%% file: figures/network_solar_cycles.pdf_tex
%% Creator: Inkscape inkscape 0.92.5, www.inkscape.org
%% Accompanies image file 'network_solar_cycles.pdf' (pdf, eps, ps)
\begingroup%
  \makeatletter%
  \providecommand\color[2][]{%
    \errmessage{(Inkscape) Color is used for the text in Inkscape, but the package 'color.sty' is not loaded}%
    \renewcommand\color[2][]{}%
  }%
  \providecommand\transparent[1]{%
    \errmessage{(Inkscape) Transparency is used (non-zero) for the text in Inkscape, but the package 'transparent.sty' is not loaded}%
    \renewcommand\transparent[1]{}%
  }%
  \providecommand\rotatebox[2]{#2}%
  \newcommand*\fsize{\dimexpr\f@size pt\relax}%
  \newcommand*\lineheight[1]{\fontsize{\fsize}{#1\fsize}\selectfont}%
  \ifx\svgwidth\undefined%
    \setlength{\unitlength}{157.38824439bp}%
    \ifx\svgscale\undefined%
      \relax%
    \else%
      \setlength{\unitlength}{\unitlength * \real{\svgscale}}%
    \fi%
  \else%
    \setlength{\unitlength}{\svgwidth}%
  \fi%
  \global\let\svgwidth\undefined%
  \global\let\svgscale\undefined%
  \makeatother%
  \begin{picture}(1,0.4481738)%
    \lineheight{1}%
    \setlength\tabcolsep{0pt}%
    \put(-0.00232114,0.07288297){\color[rgb]{0,0,0}\makebox(0,0)[lt]{\lineheight{0}\smash{\begin{tabular}[t]{l}0\end{tabular}}}}%
    \put(0.44157875,0.07275755){\color[rgb]{0,0,0}\makebox(0,0)[lt]{\lineheight{0}\smash{\begin{tabular}[t]{l}1\end{tabular}}}}%
    \put(0.21598741,0.41757427){\color[rgb]{0,0,0}\makebox(0,0)[lt]{\lineheight{0}\smash{\begin{tabular}[t]{l}2\end{tabular}}}}%
    \put(0,0){\includegraphics[width=\unitlength,page=1]{network_solar_cycles.pdf}}%
    \put(0.18936234,0.02760954){\color[rgb]{0.1254902,0.29019608,0.52941176}\makebox(0,0)[lt]{\lineheight{0}\smash{\begin{tabular}[t]{l}+1\end{tabular}}}}%
    \put(0.36748852,0.29184379){\color[rgb]{1,0.79607843,0}\makebox(0,0)[lt]{\lineheight{0}\smash{\begin{tabular}[t]{l}+3\end{tabular}}}}%
    \put(0.03818495,0.26699747){\color[rgb]{0.64313725,0,0}\makebox(0,0)[lt]{\lineheight{0}\smash{\begin{tabular}[t]{l}-2\end{tabular}}}}%
    \put(0.5321783,0.07288297){\color[rgb]{0,0,0}\makebox(0,0)[lt]{\lineheight{0}\smash{\begin{tabular}[t]{l}0\end{tabular}}}}%
    \put(0.97607821,0.07275755){\color[rgb]{0,0,0}\makebox(0,0)[lt]{\lineheight{0}\smash{\begin{tabular}[t]{l}1\end{tabular}}}}%
    \put(0.75048683,0.41757427){\color[rgb]{0,0,0}\makebox(0,0)[lt]{\lineheight{0}\smash{\begin{tabular}[t]{l}2\end{tabular}}}}%
    \put(0,0){\includegraphics[width=\unitlength,page=2]{network_solar_cycles.pdf}}%
    \put(0.90198795,0.29184379){\color[rgb]{1,0.79607843,0}\makebox(0,0)[lt]{\lineheight{0}\smash{\begin{tabular}[t]{l}+3\end{tabular}}}}%
    \put(0.81655243,0.22999864){\color[rgb]{0.64313725,0,0}\makebox(0,0)[lt]{\lineheight{0}\smash{\begin{tabular}[t]{l}-4\end{tabular}}}}%
  \end{picture}%
\endgroup%